\documentclass[fleqn,usenatbib]{mnras}

\usepackage{mathptmx}

\usepackage[T1]{fontenc}

\DeclareRobustCommand{\VAN}[3]{#2}
\let\VANthebibliography\thebibliography
\def\thebibliography{\DeclareRobustCommand{\VAN}[3]{##3}\VANthebibliography}

\usepackage{graphicx}	
\usepackage{amsmath}	

\newcommand{\kev}{keV}

\newcommand{\nicer}{\textit{NICER}}

\newcommand{\xmm}{\textit{XMM-Newton}}

\newcommand{\rexcor}{\textsc{reXcor}}
\newcommand{\lobs}{$\lambda_{\mathrm{obs}}$}

\title[The Warm Corona Model in Type I AGNs]{Unveiling Energy Pathways in AGN Accretion Flows with the Warm Corona Model for the Soft Excess}

\author[D. R. Ballantyne et al.]{D. R. Ballantyne,$^{1}$\thanks{E-mail: david.ballantyne@physics.gatech.edu}
V. Sudhakar,$^{1}$ 
D. Fairfax,$^{1,2,3}$ S. Bianchi,$^{4}$ B. Czerny,$^{5}$ A. De Rosa,$^{6}$ B. De Marco,$^{7}$ \newauthor R. Middei,$^{8,9}$ B. Palit,$^{10}$
P.-O. Petrucci,$^{11}$ A. R{\'o}{\.z}a{\'n}ska$^{10}$ and F. Ursini$^{4}$
\\
$^{1}$Center for Relativistic Astrophysics, School of Physics, Georgia Institute of Technology, 837 State Street, Atlanta, GA 30332-0430, USA\\
$^{2}$Department of Physics, Morehouse College, 830 Westview Dr SW
Atlanta, GA 30314 USA\\
$^{3}$Department of Aerospace Engineering, University of Michigan, 1320 Beal Avenue, Ann Arbor, MI 48109 USA\\
$^{4}$Dipartimento di Matematica e Fisica, Universit\`{a} degli Studi Roma Tre, via della Vasca Navale 84, 00146 Roma, Italy\\
$^{5}$Center for Theoretical Physics, Polish Academy of Sciences, Al. Lotnikov 32/46, 02-668, Warsaw, Poland\\
$^{6}$INAF – Istituto di Astrofisica e Planetologie Spaziali, Via Fosso del Cavaliere, I-00133 Roma, Italy\\
$^{7}$Departament de Fìsica, EEBE, Universitat Polit\`{e}cnica de Catalunya, Av. Eduard Maristany 16, E-08019 Barcelona, Spain\\
$^{8}$INAF - Osservatorio Astronomico di Roma, Via Frascati 33, I-00040 Monte Porzio Catone, Italy\\
$^{9}$Space Science Data Center, SSDC, ASI, Via del Politecnico snc, I-00133 Roma, Italy\\
$^{10}$Nicolaus Copernicus Astronomical Center, Polish Academy of Sciences, Bartycka 18, PL-00-716 Warszawa, Poland\\
$^{11}$Univ. Grenoble Alpes, CNRS, IPAG, 38000 Grenoble, France\\
}

\date{Accepted XXX. Received YYY; in original form ZZZ}

\pubyear{2023}

\begin{document}
\label{firstpage}
\pagerange{\pageref{firstpage}--\pageref{lastpage}}
\maketitle

\begin{abstract}
The soft excess in active galactic nuclei (AGNs) may arise through
a combination of relativistic reflection and the effects of a warm
corona at the surface of the accretion disc. Detailed examination of
the soft excess can therefore constrain models of the transport and
dissipation of accretion energy. Here, we analyze 34 \xmm\
observations from 14 Type I AGNs with the \rexcor\ spectral model
which self-consistently combines emission from a warm corona with
relativistic reflection assuming a lamppost corona. The model divides
accretion energy between the disc, the warm corona, and the lamppost. The \xmm\ observations
span a factor of 188 in Eddington ratio (\lobs) and 350 in black hole
mass, and we find that a warm corona is a significant contributor to
the soft excess for 13 of the 14 AGNs with a mean warm corona heating
fraction of $0.51$. The \rexcor\ fits reveal that the fraction of
accretion energy dissipated in the lamppost is anti-correlated with
\lobs. In contrast, the relationship between \lobs\ and both the
optical depth and heating fraction of the warm corona appears to
transition from an anti-correlation to a correlation at
$\lambda_{\mathrm{obs,t}} \approx 0.15$. Therefore, at least one other
physical process in addition to the accretion rate is needed to
explain the evolution of the warm corona. Overall, we find that a warm
corona appears to be a crucial depository of accretion energy in AGNs
across a broad range of \lobs\ and black hole mass.
\end{abstract}




\section{Introduction}
\label{sect:intro}
Active Galactic Nuclei (AGNs) generate their luminosity from the release of gravitational potential energy as matter moves radially through an accretion disc. If the accretion flow is radiatively efficient then the released energy is locally converted into photons and the disc shines brightly over a wide range of wavelengths, with the peak luminosity occurring in the ultraviolet (UV) for most AGNs \citep[e.g.,][]{ss73,pringle81,fkr02}. In addition to the optically-thick thermal emission from the disc, AGNs are also strong X-ray emitters with, on average, $\la 20$\% of their bolometric luminosity released as a power-law at photon energies $\ga 1$~keV before rolling over at energies $\ga 100$~\kev\ \citep[e.g.,][]{mdp93,vf07,vf09,ricci17b,duras20}. This X-ray power-law is well described as originating in an optically-thin Comptonizing corona of hot electrons situated close to the disc and central black hole \citep[e.g.,][]{grv79,hm91,hm93,petrucci01,fab15,fab17,kara16,alston20}. While the details of how and why accretion energy flows from the disc into the corona are not fully understood, it appears that magnetic fields in the disc and corona play a crucial role \citep[e.g.,][]{mf01a,jiang19,gr20,gr23,sbd23}.

The X-ray spectrum of AGNs contains more complexity than just a simple power-law. Superimposed on the power-law are spectral lines and high-energy curvature consistent with reprocessing of the power-law in optically-thick material lying out of the line-of-sight, a process generally referred to as X-ray reflection \citep[e.g.,][]{fr10,gk10}. Moreover, most AGNs show an excess of emission above the extrapolated power-law at energies $\la 2$~\kev\ (a `soft excess'; e.g., \citealt{piconcelli2005,winter12,ricci17b}). The soft excess may arise from X-ray reflection \citep[e.g.,][]{crummy06}, but it is also possible that a warm corona located near the surface of the accretion disc can contribute all or part of the soft excess \citep[e.g.,][]{mag98,czerny03,roz15,kd18,middei18,middei19,middei20,porquet18,ursini18,ursini20,xxu21}. Similar to the hot corona that produces the X-ray power-law, a warm corona would be heated by accretion processes and Comptonize the thermal emission from the bulk of the disc. However, the warm corona is optically thick (with $\tau \ga 10$), and $\approx 100\times$ cooler than the hot corona located near to the disc \citep[e.g.,][]{petrucci18}. If it exists, the warm corona would be another location to which accretion energy liberated in the disc is transported and dissipated. Therefore, it is important to determine the properties of any warm coronae in AGNs in order to fully understand the flow of energy in an accretion disc.

As the surface of the accretion disc is irradiated by the X-ray power-law, the warm corona properties, and the resulting emission spectrum from the disc, are determined by the combined influence of both internal heating in the warm corona and the photoionization effects of the external X-rays \citep{ball20,bx20,petrucci20}. In particular, the fraction of accretion energy dissipated in the warm corona has a strong impact on the resulting AGN X-ray spectrum and could be constrained by fitting with an appropriate model. Recently, the \rexcor\ model was made publicly available by \cite{xb22}. This spectral model assumes an optically thick accretion disc that extends to the innermost stable circular orbit (ISCO) and combines the effects of a warm corona with ionized relativistic reflection to predict the X-ray spectrum from the inner region of an AGN accretion disc. By dividing the liberated accretion energy between the disc, a warm corona, and a lamppost hot corona\footnote{This differs from the \citet{petrucci20} model which divides energy between the disc and the warm corona.}, \rexcor\ provides an opportunity to quantitatively measure the fraction of accretion energy dissipated in both the warm and hot coronae. In this paper, we apply the \rexcor\ model to a sample of 34 \xmm\ observations of 14 Type I AGNs that span a factor of 188 in Eddington ratio and two decades in black hole mass. By examining how the \rexcor\ parameters vary across this sample, we will be able to better understand the nature of the warm corona in AGNs and how accretion energy is distributed through the disc. 

The next section provides a short review of the assumptions and parameters of the \rexcor\ model, and the properties of the \xmm\ AGN sample are described in Sect.~\ref{sect:sample}. The results of fitting \rexcor\ models to the AGN sample are presented in Sect.~\ref{sect:res} 
The implications of our results on our understanding of the AGN soft excess and the warm corona model are discussed in Sect.~\ref{sect:discuss}. Finally, we summarize and draw conclusions in Sect.~\ref{sect:concl}.

\section{Review of the \rexcor\ AGN Spectral Model}
\label{sect:rexcor}
The \rexcor\ model is fully described by \cite{xb22} and so only a brief review of its features is presented here. A \rexcor\ spectrum is calculated from the inner $400$~$r_g$ (where $r_g=GM/c^2$ is the gravitational radius of a black hole with mass $M$) of an AGN accretion disc, illuminated by an X-ray power-law from a hot corona placed at a height of $h$ above the central black hole (i.e., a lamppost geometry; e.g., \citealt{matt91,mm96,dauser13}). The power-law has a photon-index $\Gamma$ and has low and high energy cutoffs at $30$~eV and $300$~\kev, respectively. The black hole has a spin $a$, and the fitting functions of \cite{fk07} are used to account for how relativistic light-bending affects the illumination of the disc. In addition to this external radiation from the hot corona, \rexcor\ also includes internal heating from a warm corona of Thomson thickness $\tau$ at the surface of the disc. This heated layer has a fixed density and interacts with the external X-rays and the thermal blackbody emission arising from the bulk of the disc. 

In \rexcor\ the luminosity of the lamppost and the amount of heat released in the warm corona are fixed fractions of the accretion disc dissipation rate $D(r, \lambda)$, where $r$ is the radius of the disc (in units of $r_g$) and $\lambda$ is the Eddington ratio \citep{ss73}. A fraction $f_X$ of $D(r,\lambda)$ within $r=10$~$r_g$ powers the hot corona luminosity, while an energy flux $h_f D(r,\lambda)$ is uniformly distributed in the warm corona layer of the disc (with Thomson depth $\tau$). Any remaining flux at a particular radius, $(1-f_X-h_f)D(r,\lambda)$, is injected into the bottom of the warm corona zone as a blackbody. 

A \rexcor\ model is computed by splitting the disc into multiple annuli and self-consistently calculating the emission from the disc surface from each annulus. The spectrum of each ring includes (i) the effects of ionized reflection due to illumination by the power-law continuum produced by the lamppost hot corona, (ii) the emission and scattering effects of the warm corona at the disc surface, and (iii) the thermal blackbody produced by the bulk of the disc.  The effects of relativistic blurring (using the \textsc{relconv\_lp} model with an inclination angle of $30$ degrees; \citealt{dauser13}) and the radial change in disc density are included for each annulus. Finally, the individual blurred reflection spectra are then integrated to produce a single spectrum which can be used to fit data between $0.2$ and $100$~\kev.

As shown by \citet{xb22}, the parameters of the \rexcor\ models impact the resulting spectra in distinct ways. The warm corona heating fraction $h_f$ and optical depth $\tau$ both strongly influence the size, shape and smoothness of the soft excess. For a given $h_f$, a larger $\tau$ will reduce the soft excess strength and smoothness as the heat released in the warm corona is dissipated in a thicker layer of gas. Similarly, for a fixed $\tau$, an increasing $h_f$ will yield a hotter, more ionized warm corona that produces a stronger soft excess that extends to higher energies. Changes to the hot corona heating fraction $f_X$ affects the \rexcor\ model by impacting the size of the soft excess relative to the power-law, as well as influencing the ionization state of the reflection features in the spectrum (analogous to the ionization parameter $\xi$ in standard reflection models; \citealt{gk10}). Finally, the photon-index $\Gamma$ changes the shape of the spectra at energies $> 2$~keV where the impact of the power-law is most important.

\citet{xb22} released eight different \rexcor\ grids that differ in the assumed lamppost height (either $5$ or $20$~$r_g$), accretion rate (either $\lambda=0.01$ or $0.1$), and black hole spin (either $a=0.9$ or $a=0.99$). Each grid contains over $20,000$ individual spectra, spanning a broad range of $\Gamma$, $\tau$, $f_X$ and $h_f$. In this paper, we make use of new \rexcor\ grids\footnote{These expanded grids are now available through the \textsc{xspec} website.} that increases the range of $\Gamma$ (from $\Gamma=1.7$--$2.2$ to $1.5$--$2.2$) with the other parameters unchanged from the original release ($0.02 \leq f_X \leq 0.2$; $0 \leq h_f \leq 0.8$; $10 \leq \tau \leq 30$).  Therefore, fitting AGN X-ray spectra with \rexcor\ grids can test for the presence of a warm corona (i.e., if $h_f >0$), and constrain how the accretion energy is divided between the lamppost and the disc surface. We note that since the \rexcor\ models assume the presence of an optically thick accretion disc down to the ISCO and a lamppost corona, then, by construction, $f_X$ is not allowed to be $0$, and \rexcor\ models always contain a contribution from relativistic ionized reflection. 

\section{Description of AGN Sample and Spectral Fitting Method}
\label{sect:sample}
\rexcor\ models are applied to a sample of \xmm\ observations of bright, radio-quiet Type I AGNs.  Our sample is drawn from the one studied by \citet{petrucci18}, which was selected to have good UV coverage with the Optical Monitor. The UV photometry is not used when fitting \rexcor\ models, so this criterion should not influence our results. 

From this parent sample, we select only those observations with photon indices inside the \rexcor\ range and have $>150,000$ counts in their EPIC-pn $0.3$--$12$~keV spectra. Observations with total counts below this limit did not provide useful constraints on the key \rexcor\ parameters (e.g., $h_f$, $f_X$ and $\tau$) despite acceptable fits. Three of the five observations of NGC4593 that lie above our count limit (Obs.\ IDs 0740920201, 0740920501 \& 0740920601) also give poorly constrained \rexcor\ parameters (see also \citealt{xb22}) and are removed from further analysis. Our final sample therefore contains 34 \xmm\ observations of 14 AGNs, and spans over two decades in both black hole mass and accretion rate. Table~\ref{table:sample} provides the details of the sample, including the estimated bolometric Eddington ratio from each observation, \lobs, as determined by \citet{petrucci18}.

\begin{table}
    \centering
    \caption{Details of the AGN sample, including the \xmm\ Observation ID, redshift, black hole mass (from \citealt{bianchi09}) and bolometric Eddington ratio ($\lambda_{\mathrm{obs}}$; estimated by \citealt{petrucci18}). All sources are radio-quiet Type I AGNs and have $>150,000$~counts in their $0.3$--$12$~\kev\ EPIC-pn spectra. \citet{petrucci18} provide details on the selection of the parent sample.}
    \label{table:sample}
    \begin{tabular}{l|c|c|c|c}
    Object & Redshift & $\log (M_{\mathrm{BH}}/M_{\odot})$ & Obs.\ ID & $\lambda_{\mathrm{obs}}$ \\ \hline
    1H0419-577 & $0.104$ & $8.58$ & 0604720301 & $0.156$ \\
     & & & 0604720401 & $0.128$ \\
    ESO198-G24 & $0.0455$ & $8.48$ & 0305370101 & $0.012$ \\
     & & & 0067190101 & $0.013$ \\
    HE1029-1401 & $0.0858$ & $8.73$ & 0203770101 & $0.102$ \\
    IRASF12397+3333 & $0.0435$ & $6.66$ & 0202180201 & $0.615$ \\
    MRK279 & $0.0304$ & $7.54$ & 0302480401 & $0.127$ \\
     & & & 0302480501 & $0.12$ \\
     & & & 0302480601 & $0.121$ \\
    MRK335 & $0.0257$ & $7.15$ & 0600540501 & $0.186$ \\
     & & & 0600540501 & $0.172$ \\
    MRK509 & $0.0343$ & $8.16$ & 0601390201 & $0.127$ \\
     & & & 0601390301 & $0.123$ \\
     & & & 0601390401 & $0.157$ \\
     & & & 0601390501 & $0.176$ \\
     & & & 0601390601 & $0.197$ \\
     & & & 0601390701 & $0.157$ \\
     & & & 0601390801 & $0.146$ \\
     & & & 0601390901 & $0.131$ \\
     & & & 0601391001 & $0.135$ \\
     & & & 0601391101 & $0.134$ \\
     MRK590 & $0.0263$ & $7.68$ & 0201020201 & $0.009$ \\
     NGC4593 & $0.00831$ & $6.73$ & 0109970101 & $0.053$ \\
      & & & 0059830101 & 0.075 \\
     PG0804+761 & $0.1$ & $8.24$ & 0605110101 & $0.402$ \\
     PG1116+215 & $0.1765$ & $8.53$ & 0201940101 & 0.384 \\
     & & & 0554380101 & $0.404$ \\
     & & & 0554380201 & $0.373$ \\
     & & & 0554380301 & $0.425$ \\
     Q0056-363 & $0.1641$ & $8.95$ & 0205680101 & $0.053$ \\
     RE1034+396 & $0.0424$ & $6.41$ & 0675440301 & $1.691$ \\
     UGC3973 & $0.0221$ & $7.72$ & 0400070201 & $0.034$ \\
     & & & 0400070301 & $0.025$ \\
     & & & 0400070401 & $0.028$ \\
    \end{tabular}
\end{table}

We analyze the same \xmm\ spectra as \citet{petrucci18}, and details of the data reduction are provided in that paper. For all sources we fit the $0.3$--$12$~\kev\ EPIC-pn \citep{pn01} spectra, except for MRK509 which is fit only for energies $>0.76$~\kev\ to avoid large features associated with the complex multi-component warm absorber \citep{detmers11,petrucci13,kaastra14}. The spectral fits are performed with \textsc{xspec} v.12.13.0 \citep{arn96} using $\chi^2$ statistics, and errorbars are reported using a $\Delta \chi^2=2.71$ criterion. Each observation is fit with the following model (in \textsc{xspec} notation):
\[
   \mathit{phabs}\times \mathit{WA} \times (\mathit{powerlaw} + {\mathit{\rexcor}} + \mathit{xillver} ),
\]
where \textit{phabs} is neutral Galactic absorption to the AGN
\citep{hi4pi16}, \textit{WA} is a \textsc{xstar}-derived warm absorber
grid \citep{walton13} located at the redshift of the source,
\textit{powerlaw} is the primary X-ray continuum\footnote{As with
standard reflection models, this \textit{powerlaw} component is
distinct from the one used in computing the \rexcor\ model by
\citet{xb22} and is included to account for the uncertainties in the
actual disc/corona geometry in AGNs. The values of $f_X$ measured by
the spectral fits are determined by the shape and features of the
\rexcor\ models, computed in the context of a lamppost corona
\citep{xb22}.} with photon index $\Gamma$, and \textit{xillver} is a
neutral reflection spectrum \citep[e.g.,][]{gk10} accounting for
reprocessing from material far from the black hole. The photon-index
is linked across the \textit{power-law}, \textit{xillver} and
\rexcor\ models. The free parameters of the warm absorber grid are the
column density and ionization parameter. The abundances of the
\textit{WA}, \textit{xillver} and \rexcor\ models are all fixed to
Solar. The inclination angle and cutoff energy of the \textit{xillver}
component are fixed at 30 degrees and 300~\kev, respectively, in order
to match the assumptions of \rexcor. The \textit{cflux} command is
applied to the \textit{powerlaw}, \rexcor\ and \textit{xillver} models
to determine the $0.3$--$10$~\kev\ normalizations and fluxes of each component. 

The \rexcor\ grids are separated by Eddington ratio, with a set calculated for $\lambda=0.01$ and one for $\lambda=0.1$. In addition to the increase in luminosity, the higher value of $\lambda$ corresponds to a lower disc density \citep{xb22}, as predicted for radiation-pressure dominated discs \citep[e.g.,][]{sz94}. As a result, the $\lambda=0.1$ grids predict a more highly ionized disc with weaker reflection features. Here, we use the $\lambda=0.1$ grids for all observations with $\lambda_{\mathrm{obs}}> 0.05$ and the $\lambda=0.01$ grid otherwise (Appendix~\ref{app:C} shows that our results are robust to this choice). Two of the AGN observations are estimated to have $\lambda_{\mathrm{obs}}=0.053$ (Obs.\ ID 0109970101 from NGC4593 and the one from Q0056-363) and thus we try both sets of \rexcor\ grids when fitting this particular dataset. We find that the $\lambda=0.01$ grids give the lowest $\chi^2$ for Q0056-363, while the $\lambda=0.1$ grids yield the best fit for the NGC4593 observation. 

For a given $\lambda$, there are 4 \rexcor\ grids covering the two different values of lamppost height ($h=5$ or $20$~$r_g$) and black hole spin ($a=0.9$ or $0.99$). All 4 grids are used when fitting the first observation of an AGN with the results recorded for the grid yielding the lowest $\chi^2$. Once the best fit is determined, the same value of $a$ is used for any subsequent observation of that AGN\footnote{By construction, all observations have comparably good statistics, so the value of $a$ does not depend on which observation was fit first. The only exception is NGC4593, where the 2nd observation (Obs. ID 0059830101) yielded a better constraint on $a$ than the first observation.} , but both values of $h$ continue to be tested for all observations.

\section{Results}
\label{sect:res}
Tables of the results of the spectral fitting are found in Appendix~\ref{app:A} while Appendix~\ref{app:B} contains plots showing the best fit models and residuals. 
In this section, we analyze the best-fit parameters from the \rexcor\ models and demonstrate how these change with the AGN Eddington ratio\footnote{A search for correlations with the black hole mass or the estimated physical accretion rate ($\dot{M}$) did not yield any statistically significant relationships.}. A discussion showing that our model assumptions do not impact the results in this Section is found in Appendix~\ref{app:C}.

\subsection{Basic Properties of Fits}
\label{sub:goodness}
Examination of the fit results in Table~\ref{table:bestfits} shows that our assumed model provides a good description of the spectrum for all 34 observations. The median reduced $\chi^2$ of the sample is $1.20$, and the maximum ($1.47$) and minimum ($0.98$) values indicate the model complexity is appropriate for the data quality. We emphasize that the goal of this spectral fitting exercise is not to find the best possible description of each spectrum, but to test if a straightforward model that includes a \rexcor\ component is a satisfactory description of the data. The $\chi^2$ values obtained here clearly show that this is the case for all 34 observations. 

We find that ten of the 34 observations do not require a warm absorber component, but MRK335 and NGC4593 are best fit with 2 warm absorbers, similar to the results of earlier studies \citep[e.g.,][]{longinotti13,ursini16}. Interestingly, the AGN with the largest estimated Eddington ratio (RE1034+396, $\lambda_{\mathrm{obs}}=1.691$) is the only source that does not need a \textit{xillver} component in the best fit model.

Nine of the 14 AGNs exhibit narrow residuals below 1~keV after fitting with our baseline model. In these cases, Gaussian components are added to the model to account for these residuals. These components are only added if the improvement is significant at $>99.9$\% confidence as measured by the F-test. The centroid energies of the Gaussians correspond to emission lines from highly ionized C, N and O (e.g., N~\textsc{vii} at $\approx 0.5$~keV, or N~\textsc{vi} at $\approx 0.43$~keV). For almost all observations a single Gaussian is added, but in two observations of UGC3973 we find that two emission lines improve the fit. In one case (1H0419-577), a Gaussian absorption line at $0.62$~keV (likely a blend of O~\textsc{vii} and O~\textsc{viii} lines) improves the fit rather than the addition of a second warm absorber. The fit statistic is found to be acceptable after the inclusion of these Gaussians, so further components are not added even if the residuals may indicate the presence of additional features (Fig.~\ref{fig:bestfits}). Although these Gaussians are added to account for clear residuals in the data, they span such a small range of energies that their presence does not influence the best-fitting \rexcor\ parameters.

In all cases the added Gaussians are narrow indicating they arise from more distant ionized gas than considered by \rexcor. Our spectral model only includes 2 sources of emission for photoionized gas --- one \rexcor\ and one \textit{xillver} component. Given the likely complexity of ionized gas around AGNs, it is not surprising that these two models cannot entirely account for all the emission from ionized gas in the observed spectra. Allowing for non-Solar abundances may also account for some of these residuals, although, apart from Fe, relaxing this assumption is not possible in \textit{xillver} and the abundances in the \rexcor\ models are currently fixed at Solar. Non-Solar abundances, as well as allowing changes in the inclination angle and high-energy cutoff, may also improve the fits at higher energies ($\ga 7$~keV) for several sources; however, the key \rexcor\ parameters ($h_f$ and $\tau$) are largely determined by the soft excess size and shape and so the impact of these high energy residuals will be minor.

\subsection{Relationships with Eddington Ratio}
\label{sub:eddratio}
The successful \rexcor\ fits described above give estimates for how the accretion energy in our sample of AGNs is distributed between the disc, a warm corona, and the hot, X-ray emitting corona. It is therefore interesting to consider how changes in the Eddington ratio of the AGNs will impact the flow of accretion energy dissipated in the system. We also include in this analysis the mean values of the \rexcor\ parameters found by \citet{xb22} when fitting five joint $\textit{XMM-Newton/NuSTAR}$ observations of the quasar HE1143-1820 with the $h=20$ \rexcor\ models. The statistical analysis is performed using Kendall's $\tau$ correlation coefficient as implemented in \textsc{pymccorrelation} \citep{isobe86,privon20} with results summarized in Table~\ref{table:stats}. To compute the correlation coefficient and $p$-value uncertainties we take the largest of the two error-bars on each data point and assume a symmetric Gaussian distribution about the data which is then sampled $10^4$ times. For data pegged at the upper or lower bounds of our parameter space, the single error-bar is used to define the symmetric distribution.
\begin{table*}
    \centering
    \caption{Summary of the correlation analysis performed on the \rexcor\ parameters and the observed Eddington ratio, \lobs. The top part of the Table gives Kendall's $\tau$ correlation coefficient and the corresponding $p$-value for the relationships with $f_X$ (Fig.~\ref{fig:fx}), $\Gamma$ (Fig.~\ref{fig:gamma}), and the $0.3$--$10$~keV flux ratio (Fig.~\ref{fig:fluxratio}). The lower part of the table provides the average $p$-values for different transition Eddington ratios ($\lambda_{\mathrm{obs,t}})$ when testing if the $\tau$--$\lambda_{\mathrm{obs}}$ and $h_f$-$\lambda_{\mathrm{obs}}$ relations (Fig.~\ref{fig:tauandhf}) changes from a negative to positive correlation at $\lambda_{\mathrm{obs,t}}$. The error-bars on Kendall's $\tau$ and $p$-values are computed from sampling the uncertainties in each of the \rexcor\ parameters $10^4$ times. The largest of the two plotted error-bars (or the single error-bar for pegged data) is used to define the symmetric Gaussian uncertainty for each point. All statistical calculations performed using \textsc{pymccorrelation} \citep{isobe86,privon20}.}
    \label{table:stats}
    \begin{tabular}{l|c|c}
    Relationship with \lobs\ & Kendall's $\tau$ & $p$-value \\ \hline
    $f_X$ & $-0.57^{+0.06}_{-0.05}$ & $1.6^{+16}_{-1.4}\times 10^{-6}$ \\
    $\Gamma$ & $0.26\pm 0.06$ & $0.031^{+0.063}_{-0.023}$ \\
    $(0.3-10)$~keV Flux ratio & $0.55\pm 0.03$ & $3.2^{+7.5}_{-2.2}\times 10^{-6}$ \\
    $(0.3-10)$~keV Flux ratio (no RE1034+396) & $0.53\pm 0.03$ & $1.3^{+3}_{-0.9}\times 10^{-5}$ \vspace{4mm} \\
    $\tau$ and $h_f$ transition \lobs\ (i.e., $\lambda_{\mathrm{obs,t}}$) &  Mean $p$-value & Mean $p$-value (no MRK335)\\ \hline
    $0.10$ & $0.10^{+0.24}_{-0.08}$ & $0.088^{+0.216}_{-0.067}$\\
    $0.12$ & $0.046^{+0.135}_{-0.033}$ & $0.049^{+0.111}_{-0.037}$\\
    $0.15$ & $0.0046^{+0.0149}_{-0.0034}$ & $0.0050^{+0.0186}_{-0.0039}$\\
    $0.20$ & $0.095^{+0.103}_{-0.060}$ & $0.12^{+0.086}_{-0.088}$\\
    \end{tabular}
\end{table*}

\begin{figure*}
    \includegraphics[width=0.99\textwidth]{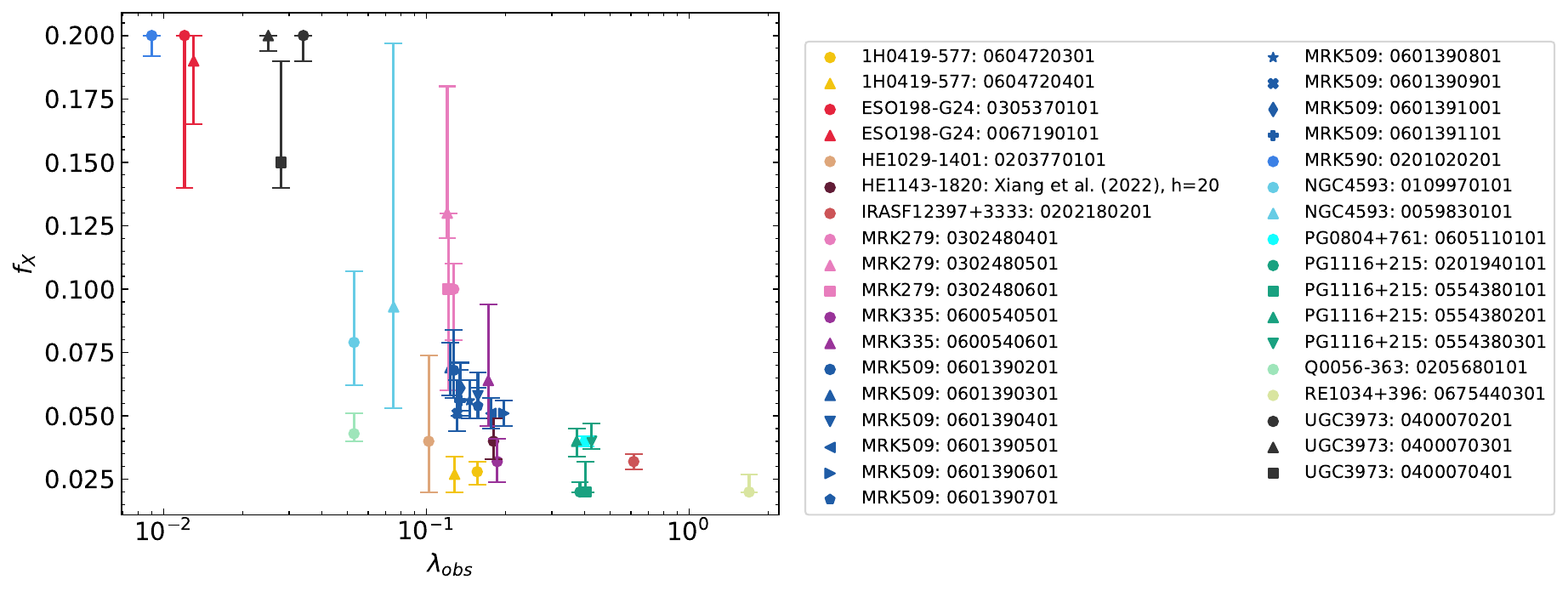}
    \caption{The hot corona heating fraction $f_X$ versus the observed Eddington ratio \lobs\ as determined from fitting the \rexcor\ spectral model to the AGN sample listed in Table~\ref{table:sample}. Each symbol on the plot shows the result from fitting one \xmm\ spectrum, and multiple observations of the same AGN have symbols with the same colour.  The $f_X$ parameter is a measure of the fraction of the accretion power dissipated in the lamppost corona. The \rexcor\ fits (given in Tables~\ref{table:bestfits}--\ref{table:WA} and shown in Fig.~\ref{fig:bestfits}) result in a clear anticorrelation between $f_X$ and \lobs, with Kendall's $\tau=-0.57^{+0.06}_{-0.05}$ (a $p$-value of $1.6^{+16}_{-1.4}\times 10^{-6}$). This result is consistent with the well established decrease in X-ray power with \lobs\ as measured by bolometric corrections \citep[e.g.,][]{vf09,duras20}.} 
    \label{fig:fx}  
\end{figure*}

Figure~\ref{fig:fx} plots $f_X$, the fraction of accretion energy dissipated in the X-ray emitting hot corona, against \lobs\ for all observations in the sample. Each AGN is shown with a different colour with individual observations separated by symbol shape. 
A clear correlation is seen in the figure with Kendall's
$\tau=-0.57^{+0.06}_{-0.05}$, which corresponds to a $p$-value of
$1.6^{+16}_{-1.4}\times 10^{-6}$ and indicates a statistically
significant anti-correlation
(Table~\ref{table:stats}){\footnote{In contrast, $f_X$ and the
  $0.3$--$10$~keV luminosity of the power-law ($\log L_{\mathrm{pl}}$)
  are only modestly correlated with Kendall's
  $\tau=-0.37^{+0.07}_{-0.06}$ and
  $p$-value$=0.0017^{+0.0093}_{-0.0015}$. Thus, the fraction of accretion
energy dissipated in the hot corona does not appear to simply
translate to the power-law luminosity, but is more dependent on the
Eddington ratio of the system.}  Physically, this relationship shows that the fraction of accretion power released in the hot, X-ray emitting corona drops from $\ga 20$\% when \lobs$\sim 0.01$ to $\la 5$\% when \lobs$\ga 0.2$ \cite[e.g.,][]{kd18}. Although the range of $f_X$ provided by \rexcor\ limits a quantitative comparison, these values and the corresponding relationship are very similar to the ones found in studies of how the X-ray bolometric correction changes with Eddington ratio \citep[e.g.,][]{vf09,duras20}. The fact that fits with the \rexcor\ model naturally leads to this relationship from our AGN sample suggests that \rexcor\ may realistically describe how the distribution of energy changes with \lobs.

The \rexcor\ fit results lead to a second relationship between the X-ray spectrum and \lobs\ that has been previously discussed in the literature. Figure~\ref{fig:gamma} shows that the X-ray photon index $\Gamma$ found by the \rexcor\ fits tends to increase with Eddington ratio \citep[e.g.,][]{brandt97,shemmer06,shemmer08,ris09,bright13,ricci13,trak17,tort23}.
\begin{figure}
\centering
    \includegraphics[width=0.5\textwidth]{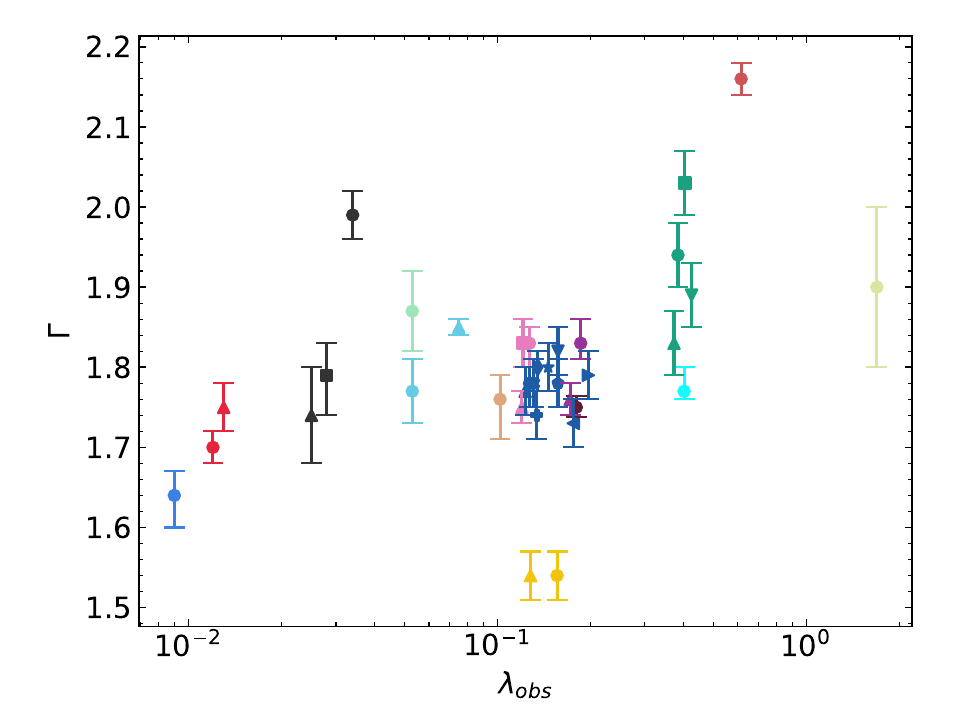}
    \caption{The relationship between the AGN photon-index $\Gamma$ and Eddington ratio \lobs\ as found from the \rexcor\ fits (Table~\ref{table:bestfits}) The symbols and colours are the same as in Fig.~\ref{fig:fx}. A moderately significant correlation is found between $\Gamma$ and \lobs\ (Kendall's $\tau=0.26\pm 0.06$, with a $p$-value of $3.1^{+6.3}_{-2.3}\times 10^{-2}$). The slope of the $\Gamma$-$\log \lambda$ correlation is $0.099\pm 0.038$, consistent with previous measurements \citep[e.g.,][]{ricci13,trak17,tort23}.}
\label{fig:gamma}
\end{figure}
As can be seen in the plot, the strength of this relationship is weaker than the $f_X$ one with Kendall's $\tau=0.26\pm 0.06$ (a $p$-value of $3.1^{+6.3}_{-2.3}\times 10^{-2}$; Table~\ref{table:stats}). The large degree of scatter seen in the Figure is similar to that found by other authors \citep[e.g.,][]{fanali13,trak17,diaz23}, and is a natural outcome of the origin of the power-law in a dynamic, Comptonizing corona \citep{ricci18}. Fitting a linear function to the data ($\Gamma=\psi \log \lambda_{\mathrm{obs}} + b$) yields $\psi=0.099\pm 0.038$ and $b=1.89\pm 0.04$, consistent with earlier measurements \citep{ricci13,trak17,tort23}. This result, along with the $f_X-\lambda_{\mathrm{obs}}$ correlation shown in Fig.~\ref{fig:fx}, together shows the applied spectral model, built around a \rexcor\ component, gives a description of the hard X-ray spectrum of AGNs in agreement with ones found by previous phenomenological models. 

Now that we have demonstrated that the fits recover prior results on the X-ray characteristics of AGNs, we consider the warm corona parameters in the \rexcor\ model and how these change with \lobs. Figure~\ref{fig:tauandhf} plots the optical depth $\tau$ and heating fraction $h_f$ of the warm corona as functions of \lobs.  
\begin{figure*}
    \includegraphics[width=0.49\textwidth]{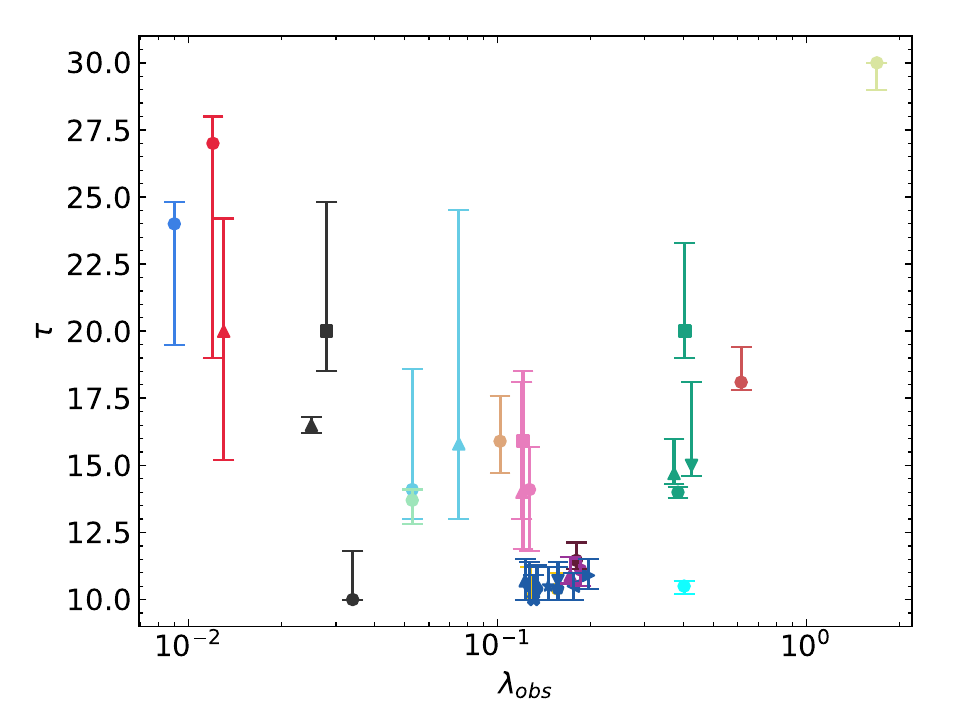}
    \includegraphics[width=0.49\textwidth]{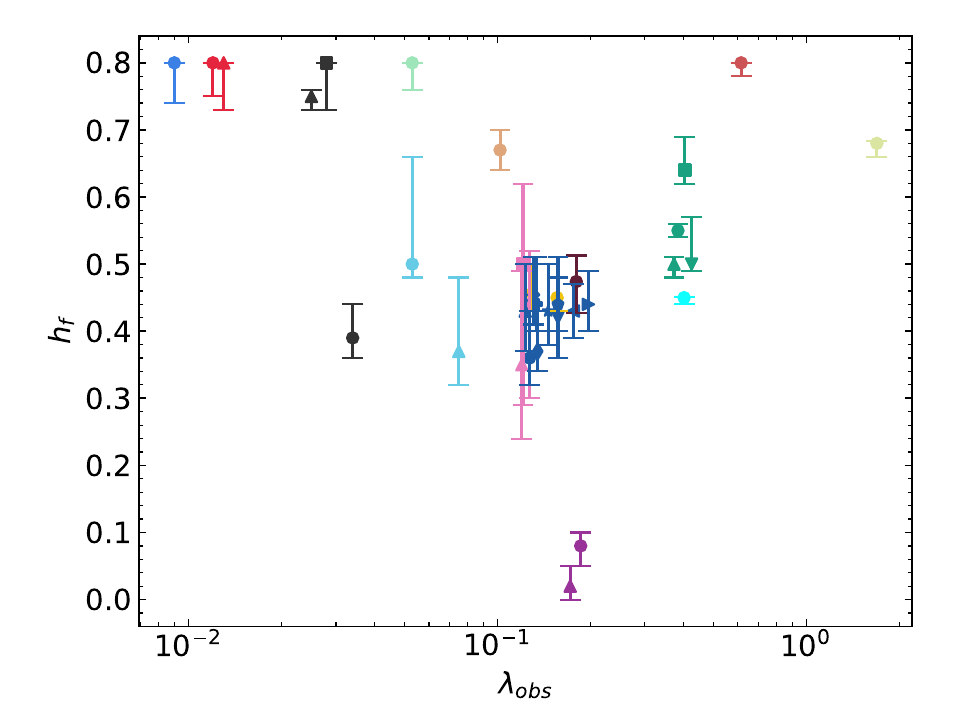}
    \caption{The left panel shows the warm corona optical depth $\tau$ found from the \rexcor\ fits plotted against the observed Eddington ratio \lobs\ of our AGN sample (Table~\ref{table:bestfits}). The right panel plots the fraction of the accretion flux dissipated in the warm corona $h_f$ versus \lobs. The symbols and colours in both panels are the same as in Fig.~\ref{fig:fx}. Aside from one observation of MRK335 (Obs. ID 0600540601), all fits give a $h_f$ significantly greater than zero, indicating that heating in the warm corona may be important for almost all AGNs in the sample. In both panels, we see evidence for a transition from an anti-correlation with \lobs\ to a correlation with \lobs\ at $\lambda_{\mathrm{obs,t}} \approx 0.15$ (Table~\ref{table:stats}).}
    \label{fig:tauandhf}
\end{figure*}
A key initial finding is that 33 out of the 34 AGN observations (in addition to the results from HE1143-1820 from \citealt{xb22}) are fit with $h_f$ significantly greater than zero with the average $h_f=0.51$. This fact suggests that a warm corona appears to be a common element of the accretion discs in the AGNs that comprise our sample, with $\approx 50$\% of the accretion power released in the warm corona. The mean optical depth of the warm corona is $\tau=14.4$, consistent with values found with fits using phenomenological models \citep[e.g.][]{jin12,petrucci13,petrucci18,yu23}.

Unlike with $f_X$ and $\Gamma$, there is clearly no correlation of $h_f$ and $\tau$ across the sampled range of \lobs. Rather, there is evidence that both parameters show a `v'-like shape that appears to transition from an anti-correlation to a correlation between $\lambda_{\mathrm{obs}}\approx 0.1$--$0.2$. This `v'-shape is most easily seen in the $\tau$-\lobs\ panel, but its presence in the $h_f$-\lobs\ plot is supported by the significant correlation we find between $h_f$ and $\tau$ (Kendall's $\tau=0.42^{+0.07}_{-0.08}$, $p$-value=$4.5^{+39}_{-4.2}\times 10^{-5}$) which suggest that $h_f$ could follow the same trend with \lobs. Based on this shape, we search for a transition $\lambda_{\mathrm{obs,t}}$ where both $h_f$ and $\tau$ are individually anti-correlated with \lobs\ below this value and correlated with \lobs\ above it. We consider $\lambda_{\mathrm{obs,t}}=0.10, 0.12, 0.15$ and $0.2$ and average the 4 $p$-values that result from Kendall's test for each value of $\lambda_{\mathrm{obs,t}}$. A clear minimum\footnote{These results are unchanged if the two MRK335 observations are removed from the analysis (Table~\ref{table:stats}).} is found with $\lambda_{\mathrm{obs,t}}=0.15$ with an average $p$-value$=4.6^{+15}_{-3.4}\times 10^{-3}$ (Table~\ref{table:stats}). The next largest average $p$-value is $10\times$ greater and occurs when $\lambda_{\mathrm{obs,t}}=0.12$. Thus, there is potentially a significant change in the relationship of the warm corona with \lobs\ at $\lambda_{\mathrm{obs,t}}=0.15$. The implications of this result is discussed in detail in Sect.~\ref{sect:discuss}.

Figure~\ref{fig:fluxratio} shows one other significant correlation that arises from the spectral fits: the \rexcor\ to power-law $0.3$--$10$~keV flux ratio increases with \lobs\ (Kendall's $\tau=0.55\pm 0.03$; $p$-value$=3.2^{+7.5}_{-2.2}\times 10^{-6}$; Table~\ref{table:stats}). Removing RE1034+396 at $\lambda_{\mathrm{obs}}=1.691$ from the sample only reduces Kendall's $\tau$ to $0.53\pm 0.03$ and the correlation remains significant ($p$-value$=1.3^{+3}_{-0.9}\times 10^{-5}$; Table~\ref{table:stats}). 
\begin{figure}
\centering
    \includegraphics[width=0.5\textwidth]{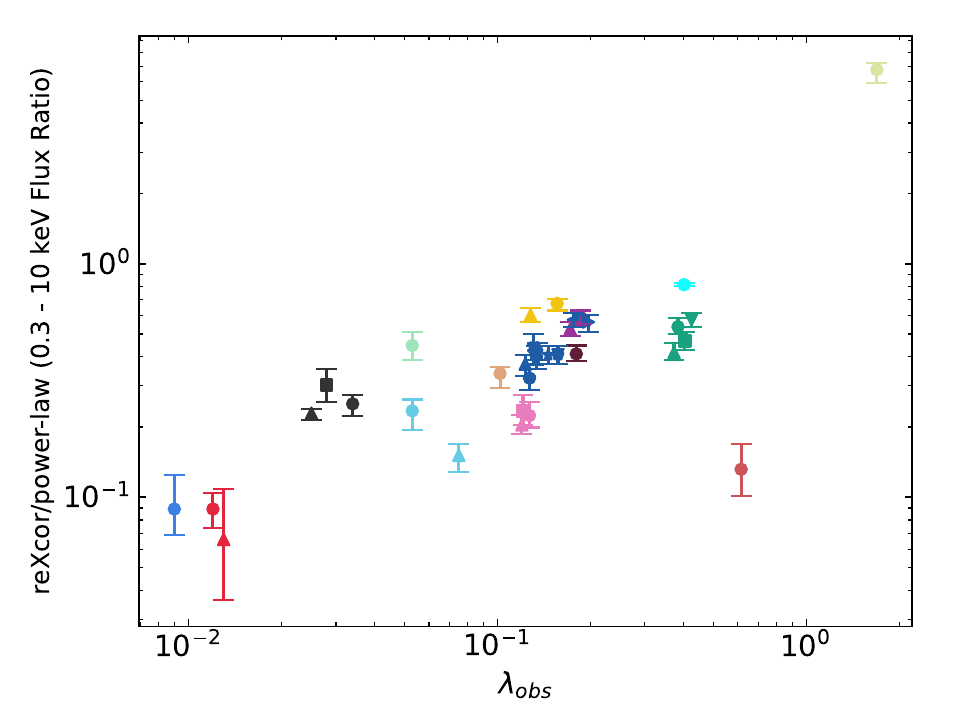}
    \caption{The ratio of the $0.3$--$10$~keV \rexcor\ and power-law fluxes obtained from the best fits (Table~\ref{table:bestfits}) as a function of \lobs. A significant correlation is found between these two quantities with a $p$-value of $3.2^{+7.5}_{-2.2}\times 10^{-6}$ (Kendall's $\tau=0.55\pm 0.03$; Table~\ref{table:stats}). The correlation persists even after removing the RE1034+396 point at $\lambda_{\mathrm{obs}}=1.691$. In this case, the $p$-value increases to $1.3^{+3}_{-0.9}\times 10^{-5}$. The increase in the \rexcor\ component relative to the intrinsic power-law is consistent with the observed strengthening of the soft excess in AGNs with higher Eddington ratios \citep[e.g.,][]{gw20,wg20,yu23}.}
    \label{fig:fluxratio}
\end{figure}
The flux ratio increases by a factor of $\approx 5$ between $\lambda_{\mathrm{obs}} \approx 0.01$ and $0.1$ and then rises more gradually. The strengthening of the \rexcor\ component relative to the power-law is consistent with the increase in the strength of the soft excess with \lobs\ observed (with significant scatter) in Seyfert galaxies \citep[e.g.,][]{gw20,wg20,yu23}. Interestingly, the rise in the relative strength of \rexcor\ happens over the entire range of \lobs\ even though $h_f$, the fraction of accretion energy released in the warm corona, reaches a minimum at $\lambda_{\mathrm{obs}} \approx 0.1$--$0.2$ (Fig.~\ref{fig:tauandhf} right). Sect.~\ref{sect:discuss} presents the interpretation of the increasing flux ratio, along with the changes in $\tau$ and $h_f$, in the context of AGN warm coronae.

\section{Discussion}
\label{sect:discuss}
The 34 \xmm\ AGN observations considered in this paper (Table~\ref{table:sample}) are all well fit with a spectral model that utilizes the \rexcor\ model to account for both the soft excess and relativistic reflection. Section~\ref{sect:res} showed the resulting relationships between the \rexcor\ parameters ($f_X$, the hard X-ray heating fraction; $\tau$, the optical depth of the warm corona; $h_f$, the heating fraction of the warm corona) and the observed Eddington ratio of the AGNs. Appendix~\ref{app:C} shows that these relationships are not strongly affected by model-dependent uncertainties and appear to be robust to changes in assumptions regarding $\lambda$, $h$ or $a$. Below, we discuss the implications of these results for understanding the role of warm coronae in AGN accretion flows.

\subsection{The Warm Corona and the Soft Excess in AGNs}
\label{sub:interp}
A basic, yet crucial, result from our analysis is that $h_f \neq 0$ for all AGNs in the sample (Table~\ref{table:bestfits}), implying that a warm corona is common in AGNs across a broad range of black hole masses and Eddington ratios (see, e.g., \citealt{porquet23}). However, since the mean $h_f=0.51$ the observed soft excess is not entirely a result of the warm corona as relativistic reflection will also contribute, and, in some cases (e.g., MRK335), even dominate the soft excess.  It is reasonnable to believe that modeling of the soft excess in X-ray observations of AGNs may require both a warm corona and relativistic reflection  \citep[e.g.,][]{porquet18,porquet21,xu21}. Our results naturally agree with this approach, as both warm corona emission and relativistic reflection are included in \rexcor\ by construction. Since both $f_X$ and $h_f$ appear to increase toward small \lobs, the fraction of accretion energy dissipated in the disc is lowest for $\lambda_{\mathrm{obs}} \sim 0.01$. The \rexcor\ results for AGNs with higher \lobs\  suggest that the accretion energy is distributed more evenly between the disc and the two coronae with the disc contributing the largest fraction at $\lambda_{\mathrm{obs}} \approx 0.1-0.2$.

 Recently, \citet{gr23} developed a one-dimensional theoretical model of an AGN warm corona that includes magnetic heating and pressure support, Compton and free-free cooling, and the effects of radiation and gas pressure. \citet{gr23} showed that a stable warm corona can develop on the surfaces of AGN accretion discs and how their properties depend on the accretion rate and magnetic properties of the disc. Specifically, the model predicts that the optical depth of the warm corona increases either with  the accretion rate or with the strength of the magnetic viscosity. The amount of energy released in the warm corona (similar to \rexcor's $h_f$) is predicted to reduce with weaker magnetic viscosity, but also be at its highest value for lower accretion rates. 

The most intriguing aspect of the spectral fitting results shown in Sect.~\ref{sub:eddratio} is that the optical depth and heating fraction of the warm corona may potentially transition from decreasing to increasing with \lobs\ at $\lambda_{\mathrm{obs,t}}\approx 0.15$. Although a direct comparison with the theoretical results of \citet{gr23} is not possible, the trends seen in Fig.~\ref{fig:tauandhf} appears to be consistent with the scenario where the strength of the magnetic viscosity in AGN discs is relatively large at low accretion rates but decreases with \lobs\ until $\lambda_{\mathrm{obs,t}} \approx 0.15$, at which point it remains unchanged or only evolves slowly. According to the model of \citet{gr23}, this would explain why $\tau$ and $h_f$ are large at low \lobs, drop to a minimum at $\lambda_{\mathrm{obs,t}} \approx 0.15$ before increasing with higher accretion rates. Stronger magnetic viscosity at low \lobs\ could also be a potential explanation for the larger values of $f_X$ in this regime (Fig.~\ref{fig:fx}) as it may lead to increasing magnetic buoyancy \citep{gr23} resulting in more efficient heating in the optically-thin, X-ray emitting corona. 

The increase in the relative strength of the \rexcor\ component with \lobs\ (Fig.~\ref{fig:fluxratio}) does not follow any of the correlations of the \rexcor\ parameters. In particular, although this correlation is consistent with the observed strengthening of the soft excess with \lobs\ \citep[e.g.,][]{gw20,wg20,yu23}, it is not related to the amount of heat dissipated in the warm corona ($h_f$). One plausible interpretation of the increase in the $0.3$--$10$~keV flux ratio is that it results from a growth in the radial size of the warm corona in the AGN accretion disc. That is, the warm corona spans a small radial range at low \lobs\ than at large \lobs. A physically smaller warm corona will produce a smaller flux compared to the power-law even if it has a large value of $h_f$ while a warm corona that takes up more area will naturally yield a larger flux compared to the power-law. The theoretical model of \citet{gr23} does predict that the size of a warm corona will increase with accretion rate, which provides some support for this interpretation.

The acceptable spectral fits and the resulting relationships with \lobs\ show that studying AGNs with models such as \rexcor\ can produce insight into the energy dissipation processes in the accretion flow. However, more theoretical work is needed in order to provide a clear understanding of the physical processes. While the \citet{gr23} model suggests a potential interpretation for the behaviors of $f_X$, $\tau$ and $h_f$, this will likely be revised as more detailed models are developed. The `v' shape suggested by Fig.~\ref{fig:tauandhf} indicates that at least one other physical parameter in addition to \lobs\ is needed to describe the warm corona heating process. The relatively high degree of scatter observed in all figures also points to a significant amount of object-to-object variability (e.g., the low $h_f$ seen from MRK335), which might be expected with such a dynamical environment as a magnetic, radiation dominated acretion disc. Future \rexcor\ studies that focus on multiple observations of a single AGN (that ideally span a wide range of \lobs) may be the best strategy for more clearly revealing the relationships affecting the warm and hot coronae in AGNs.

\subsection{Coronal height, black hole spin and MRK335}
\label{sub:mrk335}
We comment here on three other results that arise from our analysis. First, 21 of the 34 observations produce a constraint on the relative height of the lamppost corona used in the \rexcor\ model (Table~\ref{table:bestfits}). These 21 observations are from 11 individual AGNs, with 3 sources preferring a `high' coronal height (1H0419-577, MRK509 and NGC4593), while 6 are best fit with a `low' height (IRASF12397+3333, ESO198-G24, MRK335, PG0804+761, Q0056-363, and RE1034+396) and 2 AGNs have heights that appear to change between observations (PG1116+215 and UGC3973). The 6 AGNs that prefer the $h=5$ \rexcor\ models span a broad range in \lobs, consistent with the results from X-ray reverberation that a low lamppost height may be common across the span of AGN activity \citep[e.g.,][]{demarco13,kara16}. Recent X-ray polarization measurements from \textit{IXPE} suggest a more complex corona geometry in AGNs than an idealized spherical lamppost \citep[e.g.,][]{gianolli23}. Our results, therefore, may be best interpreted as favoring a geometry where the corona is compact and situated close enough to the black hole that light-bending effects impact the illumination pattern on the disk \citep[e.g.,][]{rm13,ball17}.

While the \rexcor\ fits are able to distinguish between the two values of black hole spin ($a=0.9$ and $a=0.99$) in 10 AGNs in our sample, these should not be considered as spin measurements for these sources. \citet{xb22} describe a degeneracy between $f_X$ and spin in the \rexcor\ models, so it is challenging to to use \rexcor\ fits to constrain spin unless the spectrum is strongly affected by relativistic blurring effects. In that case, the difference between the $a=0.9$ and $0.99$ models may be sufficient to distinguish between the two spin values. MRK335 is the only AGN in our sample where such a measurement is plausible as $h_f$ is small enough (Table~\ref{table:bestfits}) that the full spectrum, including the soft excess, is predicted to be dominated by relativistic reflection. We find that the \rexcor\ fits significantly prefer the $a=0.9$ models over the $a=0.99$ ones for MRK335 (the average $\Delta \chi^2=+19$ when using the $a=0.99$ \rexcor\ model). \citet{kb16} also found a black hole spin of $a \approx 0.9$ for MRK335 via multi-epoch spectral fitting of 5 \xmm\ observations using data only above 3~keV, providing some support for the value of the spin.

Lastly, it is interesting to consider why the two MRK335 observations have the lowest values of $h_f$ (indicating a weak, but not absent, warm corona) in the entire sample. These two observations occurred when the AGN was in an "intermediate" flux state after being in a short-lived "low" flux state, and were the first X-ray observations of the source to find a multi-component outflowing warm absorber \citep{longinotti13}. It is possible that the accretion disc in MRK335 was in a transitory state during these observations as the AGN evolved back to its historical flux level. It is also likely that warm corona properties may exhibit significant object-to-object variability. A future \rexcor\ analysis of archival MRK335 observations would be able to more ably capture the evolution of the warm corona in this source. 

Recently, \citet{zm23} estimated that the size of the region producing the soft excess in MRK335 is $\sim 2$--$4\times$ larger than the hot corona emission region. Although those authors were analyzing more recent \nicer\ data of MRK335, their estimated size is consistent with the assumed \rexcor\ geometry (i.e., a compact hot corona with a radially extended warm corona). While the $h_f$ values are found to be small in these two MRK335 observations analyzed here, the observed flux ratios are well in the middle of the distribution (Fig.~\ref{fig:fluxratio}). These results support the idea that the size of the warm corona is not strictly determined by the instantaneous energy flux dissipated within the layer, but is rather a function of the overall accretion properties of the system. 

\subsection{Caveats}
\label{sub:caveats}
The results presented and discussed above are subject to the assumptions and limitations of the \rexcor\ spectral model. In particular, \rexcor\ assumes a straightforward lamppost corona that is now disfavoured by \textit{IXPE} observations \citep[e.g.,][]{gianolli23}, and a constant density warm corona with a fixed heating fraction. The spectral model is limited to only fitting X-ray data. By construction, the accretion disc must always power the lamppost corona and the disc always extends to the ISCO, so $f_X$ can never equal $0$ and $h_f$ cannot take arbitrarily large values. In practice, this means that while we can test for the case where the soft excess is entirely the result of relativistic reflection (i.e., $h_f=0$), we are unable to test for a scenario where there is no relativistic reflection and the warm corona produces the entire soft excess. The \rexcor\ grids are calculated for only two values of $\lambda$ while the sample spans a range in \lobs\ of 188. These caveats must be kept in mind when considering any of our quantitative results (e.g., the mean $h_f=0.51$). We hope that this work will spur development of new accretion disc-corona models that can relax some of these assumptions. 

\section{Conclusions}
\label{sect:concl}
Unravelling the origin of the soft excess in AGNs is an opportunity to increase our understanding of accretion disc physics using only broadband X-ray spectroscopy. As the soft excess can be explained by a mix of relativistic reflection and emission from a warm corona, then studying how this combination changes across a broad sample of AGNs should give clues on the flow of accretion energy in the inner disc. In this paper, we report the results of such an experiment. We apply the \rexcor\ spectral model, which self-consistently includes the effects of both reflection and a warm corona \citep{xb22}, to a sample of 34 \xmm\ observations of 14 Type I AGNs that span a range of 188 in Eddington ratio and 350 in black hole mass. The fit parameters of the \rexcor\ model ($h_f$, the heating fraction of the warm corona; $\tau$, the optical depth of the warm corona; $f_X$, the heating fraction of the lamppost corona) have the potential to provide significant insights into the energetics of coronal heating in AGNs. 

The key findings of this experiment are
\begin{itemize}
    \item A basic spectral model that combines \rexcor, a \textit{power-law} with cutoffs at low and high energy, and a \textit{xillver} continuum (for distant reflection), all potentially modified by a warm absorber, gives acceptable fits to all 34 observations (with reduced $\chi^2$ between $0.98$ and $1.47$; Table~\ref{table:bestfits}-\ref{table:WA} and Appendix~\ref{app:B}). All fits assume Solar abundances, a high-energy cutoff energy of $300$~keV and an inclination angle of $30$~degrees. The warm corona parameters are largely determined by the soft excess size and shape and will therefore be minimally impacted by these assumptions.
    \item A significant anti-correlation is found between $f_X$ and \lobs, the estimated Eddington ratio of the AGNs (Fig.~\ref{fig:fx}), indicating that the fraction of accretion energy dissipated in the X-ray emitting corona falls with \lobs. This is consistent with the observed changes in the X-ray bolometric correction \citep[e.g.,][]{duras20}.
    \item The photon index and \lobs\ are correlated with moderate significance (Fig.~\ref{fig:gamma}). The slope and intercept of the relationship is consistent with previous measurements \citep[e.g.,][]{tort23}.
    \item The average value of $h_f$ in the sample is $0.51$, which shows that warm corona heating is important for modeling the soft excess in nearly all AGNs in the sample. Only the two MRK335 observations have $h_f < 0.1$, indicating that relativistic reflection dominates the soft excess in those observations. Thus, our analysis explicitly demonstrates that the soft excess in AGNs can be successfully modeled by combining the effects of a warm corona with relativistic reflection.
    \item The optical depth $\tau$ and heating fraction $h_f$ of the warm corona both show evidence for a `v'-like relationship with \lobs, transitioning from an apparent anti-correlation to a positive one at $\lambda_{\mathrm{obs,t}}\approx 0.15$ (Fig.~\ref{fig:tauandhf} and Table~\ref{table:stats}). This result suggests that at least one other physical property, in addition to the Eddington ratio, is important in determining the properties of the warm corona, e.g., the strength of the magnetic viscosity in the disc (Sect.~\ref{sub:interp}; \citealt{gr23}).
    \item The flux of the \rexcor\ component increases relative to the power-law for sources with larger \lobs\ (Fig.~\ref{fig:fluxratio}), following the known trend for the soft excess to appear stronger in more rapidly accreting AGNs \citep[e.g.,][]{gw20}. This can be most readily interpreted as the radial extent of the warm corona corona increases with \lobs.
    \item The majority of the AGNs in the sample prefer \rexcor\ models with a low lamppost height ($h=5$~$r_g$), rather than ones that assume $h=20$~$r_g$ (Sect.~\ref{sub:mrk335}), indicating that most AGN coronae are compact and lie close to the black hole, regardless of the exact geometry of the corona.
\end{itemize}

These findings provide strong evidence that both relativistic reflection and a warm corona can accurately describe the soft excess across a range of AGNs. In addition, the \rexcor\ parameters provide a means by which to study how the accretion energy is distributed in accretion flows. Future theoretical work on warm coronae is needed to explain the `v'-shaped relationships between the warm corona properties and the Eddington ratio, as at least one other physical property must be involved. Applying the \rexcor\ model to a single AGN with multiple high quality spectra will reduce the effects of object-to-object variability and may clarify some of the relationships presented here. Overall, our results show that careful analysis of the AGN soft excess will yield important new insights on the physics of accretion discs.

\section*{Acknowledgements}
D.\ Fairfax participated as part of the Georgia Tech Physics REU program supported by NSF grant no.\ 1852519. This research was supported by the International Space Science Institute (ISSI) in Bern, through ISSI International Team project \#514 (Warm Coronae in AGN: Observational Evidence and Physical Understanding). SB acknowledges support from the EU grant AHEAD-2020 (GA No. 871158). BDM acknowledges support via RYC2018-025950-I, PID2020-117252GB-I00, PID2022-136828NB-C44, and SGR-386/2021. AR and BP were supported by the Polish National Science
Center grants No. 2021/41/B/ST9/04110.

\section*{Data Availability}

The data underlying this article will be shared on reasonable request to the corresponding author. The \rexcor\ models are publicly available through the \textsc{xspec} website.



\bibliographystyle{mnras}
\bibliography{refs} 




\appendix

\section{Best Fit Parameters}
\label{app:A}
This Appendix presents the best fit parameters from the AGN sample. The results are presented in three tables, where Table~\ref{table:bestfits} lists the main \rexcor\ parameters as well as the overall $\chi^2$ of the fit, Table~\ref{table:fluxes} contains the fluxes of the 3 continuum components in the spectra model as well as details of any additional Gaussian emission lines added to the model, and Table~\ref{table:WA} provides the warm absorber parameters.


%
\begin{table*}
    \centering
    \caption{The parameters obtained from the \rexcor\ component of the best-fitting spectral model: the warm corona heating fraction $h_f$, the lamppost heating fraction $f_X$, the optical depth of the warm corona $\tau$, and the photon-index $\Gamma$. For each AGN we test \rexcor\ models with black hole spin $a=0.9$ or $0.99$, as well as coronal heights $h=5$~$r_g$ or $20$~$r_g$ with the model giving the smallest $\chi^2$ entered in the Table. For AGNs with more than one observation, the spin that leads to the lowest $\chi^2$ for the first observation is used for any subsequent observations. A `$\dag$' (`$*$') symbol by the Observation ID indicates that the fit with the other value of $h$ ($a$) has a $\Delta \chi^2 < 6$ compared to the tabulated model. A 'p' in the errorbar denotes a parameter pegging at the boundary of the \rexcor\ grid. All observations are fit between $0.3$ and $12$~keV, except for MRK509 which is fit between $0.76$ and $12$~keV to avoid the complex warm absorber features at lower energies.  }
    \label{table:bestfits}
    \begin{tabular}{l|l|c|c|c|c|c|c|c}
 Object & Obs.\ ID & $a$ & $h$ ($r_g$) & $h_f$ & $f_X$ & $\tau$ & $\Gamma$ & $\chi^2/$dof \\ \hline
1H0419-577 & 0604720301 & $0.99$ & $20$ & $0.45^{+0.03}_{-0.02}$ & $0.028^{+0.004}_{-0.005}$ & $10.4^{+0.6}_{-0.4p}$ & $1.54\pm 0.03$ & $332/266$ \\ \vspace{2mm}
 & 0604720401 & $0.99$ & $20$ & $0.46^{+0.04}_{-0.03}$ & $0.027\pm 0.007$ & $10.2^{+1.0}_{-0.2p}$ & $1.54\pm 0.03$ & $253/259$ \\
ESO198-G24 & 0305370101 & $0.99$ & $5$ & $0.80_{-0.05}^{+0p}$ & $0.20_{-0.06}^{+0p}$ & $27.0_{-8.0}^{+1.0}$ & $1.70\pm 0.02$ & $282/271$ \\ \vspace{2mm}
 & 0067190101$^{\dag}$ & & $5$ & $0.80_{-0.07}^{+0p}$ & $0.19_{-0.025}^{+0.01p}$ & $20.0_{-4.8}^{+4.2}$ & $1.75\pm 0.03$ & $331/252$ \\ \vspace{2mm}
HE1029-1401	& 0203770101$^{\dag *}$ & $0.99$ & $20$ & $0.67\pm 0.03$ & $0.040_{-0.020p}^{+0.034}$ & $15.9_{-1.2}^{+1.7}$ & $ 1.76_{-0.05}^{+0.03}$ & $313/266$ \\ \vspace{2mm}
IRASF12397+3333 & 0202180201$^{*}$ & $0.99$ & $5$ & $0.80^{+0p}_{-0.02}$ & $0.032\pm 0.003$ & $18.1^{+1.3}_{-0.3}$ & $2.16\pm 0.02$ & $313/238$ \\
MRK279 & 0302480401$^{\dag}$ & $0.99$ & $20$ & $0.44_{-0.14}^{+0.08}$ & $0.10_{-0.02}^{+0.01}$ & $14.1_{-2.3}^{+1.6}$ & $1.83_{-0.03}^{+0.02}$ & $326/270$\\
    & 0302480501$^{\dag}$ & & $5$ & $0.35_{-0.11}^{+0.14}$ & $0.13^{+0.05}_{-0.01}$ & $14.0_{-1.0}^{+4.1}$ & $1.75\pm 0.02$ & $361/270$ \\ \vspace{2mm}
    & 0302480601$^{\dag}$ & & $20$ & $0.50_{-0.21}^{+0.12}$ & $0.10_{-0.04}^{+0.03}$ & $15.9_{-4.0}^{+2.6}$ & $1.83\pm 0.03$ & $287/264$ \\
MRK335 & 0600540501 & $0.9$ & $5$ & $0.08_{-0.03}^{+0.02}$ & $0.032_{-0.008}^{+0.009}$ & $ 11.1_{-0.6}^{+0.4}$ & $1.83_{-0.02}^{+0.03}$ & $268/247$\\ \vspace{2mm}
    & 0600540601 & & $5$ & $0.02_{-0.02}^{+0.03}$ & $0.064_{-0.018}^{+0.03}$ & $10.9_{-0.3}^{+0.7}$ & $1.76\pm 0.02$ & $343/256$\\
MRK509& 0601390201$^{*}$ & $0.99$ & $20$ & $0.36_{-0.04}^{+0.07}$ & $0.068_{-0.011}^{+0.016}$ & $10.5_{-0.5p}^{+0.9}$ & $1.78_{-0.03}^{+0.02}$ & $269/248$\\
    & 0601390301$^{*}$ & & $20$ & $0.43_{-0.06}^{+0.07}$ & $0.069_{-0.011}^{+0.010}$ & $10.7_{-0.7p}^{+0.8}$ & $1.77\pm 0.03$ & $321/250$\\
    & 0601390401$^{*}$ & & $20$ & $0.42\pm 0.06$ & $0.058\pm 0.009$ & $10.7_{-0.7p}^{+0.7}$ & $1.82\pm 0.03$ & $359/250$ \\
    & 0601390501$^{*}$ & & $20$ & $0.43\pm 0.04$ & $0.051\pm 0.006$ & $10.5_{-0.5p}^{+0.5}$ & $1.73\pm 0.03$ & $367/249$ \\
    & 0601390601$^{*}$ & & $20$ & $0.44_{-0.04}^{+0.05}$ & $0.051\pm 0.005$ & $10.9_{-0.5}^{+0.6}$ & $1.79\pm 0.03$ & $305/250$ \\
    & 0601390701$^{\dag *}$ & & $20$ & $0.44_{-0.04}^{+0.07}$ & $0.054^{+0.007}_{-0.005}$ & $10.4_{-0.4p}^{+0.8}$ & $1.78\pm 0.03$ & $355/250$ \\
    & 0601390801$^{*}$ & & $20$ & $0.43^{+0.07}_{-0.05}$ & $0.056_{-0.007}^{+0.008}$ & $10.5_{-0.5p}^{+0.7}$ & $1.80\pm 0.03$ & $326/249$ \\
    & 0601390901$^{*}$ & & $20$ & $0.45_{-0.04}^{+0.06}$ & $0.051_{-0.007}^{+0.013}$ & $10.0_{-0p}^{+0.9}$ & $1.78\pm 0.03$ & $304/248$\\
    & 0601391001$^{*}$ & & $20$ & $0.37_{-0.03}^{+0.06}$ & $0.061^{+0.010}_{-0.009}$ & $10.4_{-0.4p}^{+0.8}$ & $1.80^{+0.02}_{-0.03}$ & $275/250$ \\ \vspace{2mm}
    & 0601391101$^{*}$ & & $20$ & $0.44_{-0.04}^{+0.07}$ & $0.056_{-0.006}^{+0.012}$ & $10.4_{-0.4p}^{+0.9}$ & $1.74\pm 0.03$ & $286/247$ \\ \vspace{2mm}
MRK590 & 0201020201$^{\dag}$ & $0.99$ & $5$ & $0.80_{-0.06}^{+0p}$ & $0.20_{-0.008}^{+0p}$ & $24.0_{-4.5}^{+0.8}$ & $1.64_{-0.04}^{+0.03}$ & $256/259$\\
NGC4593 & $0109970101^{\dag *}$ & $0.9$ & $20$ & $0.50^{+0.16}_{-0.02}$ & $0.079^{+0.028}_{-0.017}$ & $14.1^{+4.5}_{-1.1}$ & $1.77\pm 0.04$ & $286/254$ \\ \vspace{2mm} 
 & $0059830101^{*}$ & & $20$ & $0.37^{+0.11}_{-0.05}$ & $0.093^{+0.104}_{-0.040}$ & $15.8^{+8.7}_{-2.8}$ & $1.85\pm 0.01$ & $332/268$ \\ \vspace{2mm}
PG0804+761 & 0605110101 & $0.99$ & $5$ & $0.45^{+0.001}_{-0.01}$ & $0.040\pm 0.002$ & $10.5^{+0.2}_{-0.3}$ & $1.77^{+0.03}_{-0.01}$ & $260/246$ \\
PG1116+215 & 0201940101 & $0.99$ & $20$ & $0.55\pm 0.01$ & $0.020_{-0p}^{+0.004}$ & $14.0\pm 0.2$ & $1.94\pm 0.04$ & $302/260$ \\
 & 0554380101$^{\dag}$ & & $20$ & $0.64^{+0.05}_{-0.02}$ & $0.020^{+0.012}_{-0p}$ & $20.0^{+3.3}_{-1.0}$ & $2.03\pm 0.04$ & $326/246$ \\
& 0554380201$^{\dag}$ & & $5$ & $0.50_{-0.02}^{+0.01}$ & $0.040^{+0.005}_{-0.006}$ & $14.7^{+1.3}_{-0.4}$ & $1.83\pm 0.04$ & $284/248$ \\ \vspace{2mm}
 & 0554380301$^{\dag}$ & & $5$ & $0.50^{+0.07}_{-0.01}$ & $0.040^{+0.007}_{-0.003}$ & $15.0^{+3.1}_{-0.4}$ & $1.89\pm 0.04$ & $302/237$ \\ \vspace{2mm}
Q0056-363 & 0205680101 & $0.99$ & $5$ & $0.80^{+0p}_{-0.04}$ & $0.043^{+0.008}_{-0.003}$ & $13.7^{+0.4}_{-0.9}$ & $1.87\pm 0.05$ & $300/242$ \\ \vspace{2mm}
RE1034+396 & 0675440301 & $0.99$ & $5$ & $0.68^{+0.004}_{-0.02}$ & $0.020^{+0.007}_{-0p}$ & $30^{+0p}_{-1}$ & $1.9\pm 0.1$ & $206/143$ \\
UGC3973	& 0400070201 & $0.99$ & $5$ & $0.39_{-0.03}^{+0.05}$ & $0.20_{-0.010}^{+0p}$ & $10.0_{-0p}^{+1.8}$ & $1.99\pm 0.03$ & $286/255$\\
    & 0400070301$^{\dag}$ &  & $5$ & $0.75_{-0.02}^{+0.01}$ & $0.20_{-0.006}^{+0p}$ & $16.5\pm 0.3$ & $1.74\pm 0.06$ & $270/246$\\
    & 0400070401$^{\dag}$ &  & $20$ & $0.80_{-0.07}^{+0p}$ & $0.15_{-0.01}^{+0.04}$ & $20.0_{-1.5}^{+4.8}$ & $1.79_{-0.05}^{+0.04}$ & $248/244$\\
 \end{tabular}
 \end{table*}

\begin{table*}
    \centering
    \caption{In this table we provide the $0.3$--$10$~\kev\ fluxes ($F$, in erg~cm$^{-2}$~s$^{-1}$) of the \rexcor, \textit{power-law} and \textit{xillver} components from the best-fit models in Table~\ref{table:bestfits}. The fluxes are determined using the `cflux' command in \textsc{xspec}. RE1034+396 did not require a \textit{xillver} contribution to the best-fit model. In addition, the energy ($E$, in keV) and normalization ($K$, in ph~cm$^{-2}$~s$^{-1}$) of any narrow ($\sigma=0$) Gaussian emission lines added to the model are also tabulated. Most sources only required one line, but UGC3973 required an additional line. These lines correspond to emission features from highly ionized C, N, and O. Empty entries are denoted with `--'.}  
    \label{table:fluxes}
    \begin{tabular}{l|l|c|c|c|c|c|c}
 Object & Obs.\ ID & $\log F$ (\rexcor) & $\log F$ (\textit{power-law}) & $\log F$ (\textit{xillver}) & $E$ (keV) & $K$ ($\times 10^{-5}$) \\ 
 \hline
 1H0419-577 & 0604720301 & $-10.88^{+0.01}_{-0.02}$ & $-10.71\pm 0.01$ & $<-12.75$ & $0.62\pm 0.01$ & $-27\pm 4$ \\ \vspace{2mm}
  & 0604720401 & $-10.95\pm 0.02$ & $-10.73\pm 0.01$ & $< -12.78$ & $0.62\pm 0.01$ & $-19^{+6}_{-7}$ \\
ESO198-G24 & 0305370101 & $-11.83_{-0.07}^{+0.06}$ & $-10.78\pm 0.01$ & $-12.27^{+0.06}_{-0.08}$ & $0.88\pm 0.02$ & $1.3\pm 0.5$ \\ \vspace{2mm}
 & 0067190101 & $ -11.82_{-0.24}^{+0.21}$ & $-10.64\pm 0.01$ & $-12.57_{-0.40}^{+0.21}$ & $0.51\pm 0.02$ & $7.8\pm 2.9$ \\ \vspace{2mm}
 HE1029-1401 & 0203770101 & $-10.97_{-0.05}^{+0.02}$ & $-10.50\pm 0.01$ & $-12.18_{-0.16}^{+0.12}$ & $0.51\pm 0.01$ & $37\pm 7$ \\ \vspace{2mm}
 IRASF12397+3333 & 0202180201 & $-11.79\pm 0.10$ & $-10.91\pm 0.01$ & $-12.69^{+0.12}_{-0.21}$ & $0.48^{+0.01}_{-0.004}$ & $16\pm 2$ \\
MRK279 & 0302480401 & $-10.98_{-0.04}^{+0.05}$ & $-10.33\pm 0.01$ & $-11.85_{-0.08}^{+0.07}$ & $0.47\pm 0.01$ & $44^{+7}_{-6}$ \\
    & 0302480501 & $-11.04_{-0.03}^{+0.04}$ & $-10.35_{-0.01}^{+0.003}$ & $-11.88\pm 0.06$ & $0.46\pm 0.01$ & $22^{+12}_{-6}$ \\ \vspace{2mm}
    & 0302480601 & $-10.99_{-0.05}^{+0.06}$ & $-10.36\pm 0.01$ & $-11.80_{-0.11}^{+0.08}$ & $0.47\pm 0.01$ & $26^{+14}_{-8}$ \\
MRK335 & 0600540501  & $-11.34\pm 0.02$ & $-11.11\pm 0.01$ & $-12.31_{-0.08}^{+0.07}$ & -- & -- \\ \vspace{2mm}
    & 0600540601 & $-11.44\pm 0.02$ & $-11.16\pm 0.01$ & $-12.47_{-0.09}^{+0.07}$ & -- & -- \\
MRK509	& 0601390201 & $-10.64^{+0.03}_{-0.04}$ & $ -10.15\pm 0.01$ & $-11.88_{-0.11}^{+0.09}$ & -- & -- \\
    & 0601390301 & $-10.58^{+0.03}_{-0.04}$ & $-10.15\pm 0.01$ & $-11.84_{-0.09}^{+0.08}$ & -- & -- \\
    & 0601390401 & $-10.48\pm 0.03$ & $-10.09\pm 0.01$ & $-11.76_{-0.09}^{+0.07}$ & -- & -- \\
    & 0601390501 & $-10.46\pm 0.02$ & $-10.22\pm 0.01$ & $-11.97_{-0.12}^{+0.09}$ & -- & -- \\
    & 0601390601 & $-10.38^{+0.02}_{-0.03}$ & $-10.13\pm 0.01$ & $-11.83_{-0.10}^{+0.08}$ & -- & -- \\
    & 0601390701 & $-10.47\pm 0.03$ & $-10.08\pm 0.01$ & $-11.82_{-0.09}^{+0.08}$ & -- & -- \\
    & 0601390801 & $-10.49\pm 0.03$ & $-10.10\pm 0.01$ & $-11.78_{-0.09}^{+0.07}$ & -- & -- \\
    & 0601390901 & $-10.43^{+0.05}_{-0.06}$ & $-10.07\pm 0.01$ & $-11.78_{-0.09}^{+0.07}$ & -- & -- \\
    & 0601391001 & $-10.51_{-0.03}^{+0.02}$ & $-10.14\pm 0.01$ & $-11.79_{-0.08}^{+0.07}$ & -- & -- \\ 
\vspace{2mm}
    & 0601391101 & $-10.49\pm 0.04$ & $-10.09\pm 0.01$ & $-11.84_{-0.10}^{+0.09}$ & -- & -- \\ \vspace{2mm}
MRK590 & 0201020201 & $-12.02_{-0.10}^{+0.14}$ & $-10.97\pm 0.01$ & $-12.28_{-0.09}^{+0.08}$ & $0.86^{+0.05}_{-0.03}$ & $1.2\pm 0.5$ \\
NGC4593 & 0109970101 & $-10.80^{+0.03}_{-0.07}$ & $-10.17^{+0.02}_{-0.01}$ & $-11.64^{+0.10}_{-0.13}$ & $0.45\pm 0.01$ & $56^{+18}_{-22}$ \\ \vspace{2mm}
 & 0059830101 & $-10.97^{+0.04}_{-0.06}$ & $-10.15\pm 0.01$ & $-11.55\pm 0.05$ & $0.46\pm 0.01$ & $47\pm 13$ \\ \vspace{2mm}
PG0804+761 & 0605110101 & $-10.88^{+0.004}_{-0.003}$ & $-10.79\pm 0.004$ & $-12.39^{+0.12}_{-0.17}$ & $0.44\pm 0.01$ & $75^{+13}_{-8}$ \\
PG1116+215 & 0201940101 & $-11.47_{-0.02}^{+0.03}$ & $-11.20\pm 0.01$ & $-12.74_{-0.20}^{+0.11}$ & -- & -- \\
 & 0554380101 & $-11.35\pm 0.03$ & $-11.02\pm 0.01$ & $-12.52^{+0.14}_{-0.19}$ & -- & -- \\ 
 & 0554380201 & $-11.49^{+0.03}_{-0.02}$ & $-11.11\pm 0.01$ & $-12.69^{+0.14}_{-0.26}$ & -- & -- \\ \vspace{2mm}
 & 0554380301 & $-11.48\pm 0.02$ & $-11.24\pm 0.01$ & $-12.61^{+0.13}_{-0.07} $ & -- & -- \\ \vspace{2mm}
 Q0056-363 & 0205680101 & $-11.64\pm 0.04$ & $-11.29\pm 0.02$ & $-12.81^{+0.13}_{-0.17}$ & -- & -- \\ \vspace{2mm}
 RE1034+396 & 0675440301 &
 $-10.85_{-0.005}^{+0.01}$ & $-11.68^{+0.02}_{-0.05}$ & -- & -- & -- \\
UGC3973	& 0400070201 & $-10.96_{-0.04}^{+0.03}$ & $-10.36\pm 0.01$ & $-11.87_{-0.13}^{+0.10}$ & $0.48\pm 0.01$ & $37^{+14}_{-11}$ \\
& & & & & $0.77^{+0.01}_{-0.004}$ & $171^{+22}_{-36}$ \\
    & 0400070301 & $-11.12_{-0.02}^{+0.01}$ & $-10.48\pm 0.01$ & $-11.95_{-0.08}^{+0.07}$ & $0.77\pm 0.01$ & $43^{+6}_{-11}$ \\
    & 0400070401 & $-11.03_{-0.05}^{+0.06}$ & $-10.51^{+0.01}_{-0.02}$ & $-11.95_{-0.15}^{+0.11}$ & $0.47\pm 0.01$ & $35^{+11}_{-9}$ \\
& & & & & $0.77^{+0.02}_{-0.01}$ & $28^{+71}_{-16}$ \\

 \end{tabular}
 \end{table*}

\begin{table}
    \centering
    \caption{The column density ($N_H$ in cm$^{-2}$) and ionization parameter ($\xi$, in erg~s$^{-1}$~cm$^{-1}$) of any warm absorber in the best-fit model (Table~\ref{table:bestfits}). Observations that did not require a warm absorber are omitted from the Table. MRK335 and NGC4593 both required two warm absorbers. A 'p' in the errorbar denotes a parameter pegging at the boundary of the warm absorber grid.}  
    \label{table:WA}
    \begin{tabular}{l|l|c|c}
 Object & Obs.\ ID & $N_H$ & $\log \xi$ \\ 
 \hline
 1H0419-577 & 0604720301 & $4.6^{+2}_{-1.4}\times 10^{20}$ & $0.85^{+0.29}_{-0.49}$ \\ \vspace{2mm}
  & 0604720401 & $3.8^{+3.6}_{-1.2}\times 10^{20}$ & $0.73^{+0.49}_{-0.73p}$ \\ \vspace{2mm}
IRASF12397+3333 & 020218201 & $6.1^{+0.8}_{-0.6}\times 10^{21}$ & $1.83\pm 0.02$ \\ 
MRK279 & 0302480401 & $2.9_{-1.1}^{+1.0}\times 10^{20}$ & $1.68_{-0.36}^{+0.08}$\\
    & 0302480501 & $1.8_{-0.5}^{+0.6}\times 10^{20}$ & $0_{-0p}^{+0.37}$\\ \vspace{2mm}
    & 0302480601 & $1.5_{-1.5p}^{+0.4}\times 10^{20}$ & $0.05_{-0.05p}^{+0.5}$\\
MRK335 & 0600540501 & $4.8^{+0.8}_{-0.9}\times 10^{21}$ & $1.89\pm 0.03$ \\
 & & $1.1^{+3.6}_{-0.8}\times 10^{22}$ & $2.91^{+0.25}_{-0.17}$ \\
    & 0600540601 & $1.1^{+0.1}_{-0.1}\times 10^{22}$  & $1.87^{+0.03}_{-0.01}$ \\ \vspace{2mm}
    & & $1.3^{+1}_{-0.9}\times 10^{22}$ & $2.86^{+0.06}_{-0.13}$ \\
MRK509 & 0601390901 & $1.1_{-0.7}^{+1.0}\times 10^{21}$ & $2.03_{-0.08}^{+0.22}$\\ \vspace{2mm}
    & 0601391101 & $6.4_{-3.3}^{+4.0}\times 10^{20}$ & $2.26_{-0.22}^{+0.28}$\\
NGC4593 & 0109970101 & $3.5^{+4.1}_{-1.2}\times 10^{20}$ & $0.50^{+0.58}_{-0.38}$ \\
 & & $3.8^{+1.2}_{-0.7}\times 10^{21}$ & $2.06^{+0.04}_{-0.05}$ \\
 & 0059830101 & $5.4^{+1.2}_{-1.6}\times 10^{20}$ & $0.74^{+0.20}_{-0.18}$ \\ \vspace{2mm}
 & & $3.0^{+0.4}_{-0.5}\times 10^{21}$ & $2.16^{+0.05}_{-0.08}$ \\ \vspace{2mm}
PG0804+761 & 0605110101 & $1.0^{+0.3}_{-0p}\times 10^{20}$ & $0.63^{+0.21}_{-0.63p}$ \\ 
PG1116+215	& 0201940101 & $2.6\pm 0.4 \times 10^{20}$ & $0_{-0p}^{+0.13}$\\
& 0554380101 & $3.5^{+0.6}_{-0.5}\times 10^{20}$ & $0_{0p}^{+0.11}$ \\
& 0554380201 & $3.1^{+0.5}_{-0.4}\times 10^{20}$ & $0_{-0p}^{+0.37}$ \\ \vspace{2mm}
& 0554380301 & $4.1_{-0.4}^{+0.5}\times 10^{20}$ & $0^{+0.20}_{-0p}$ \\ \vspace{2mm}
Q0056-363 & 0205680101 & $1.0^{+1.0}_{-0p}\times 10^{20}$ & $0^{+0.82}_{-0p}$ \\ \vspace{2mm}
RE1034+396 & 0675440301 & $1.0^{+1.3}_{-0p}\times 10^{20}$ & $0.21^{+0.31}_{-0.21p}$ \\
UGC3973	& 0400070201 & $5.2_{-0.5}^{+0.3}\times 10^{21}$ & $1.69_{-0.04}^{+0.02}$\\
    & 0400070301 & $6.0_{-0.5}^{+0.2}\times 10^{21}$ & $1.69_{-0.03}^{+0.06}$\\
    & 0400070401 & $6.1_{-0.6}^{+0.9}\times 10^{21}$ & $1.73_{-0.03}^{+0.04}$\\
 \end{tabular}
 \end{table}

\section{Plots of Best Fits}
\label{app:B}
Figure~\ref{fig:bestfits} plots the best fitting spectrum for each observation using the model described in Sect.~\ref{sect:sample}. The upper portion of each panel shows the total spectrum (solid line), plus the contributions from the power-law (short-dashed line), the \rexcor\ model (dot-dashed line), and the \textit{xillver} distant reflector (dotted line). Any Gaussian lines are shown as dot-dot-dashed lines. The lower part of each panel plots the residuals from the best-fitting model in units of $\sigma$. The parameters of the best-fitting model are listed in Tables~\ref{table:bestfits}--\ref{table:WA}.
\begin{figure*}
\includegraphics[width=0.49\textwidth]{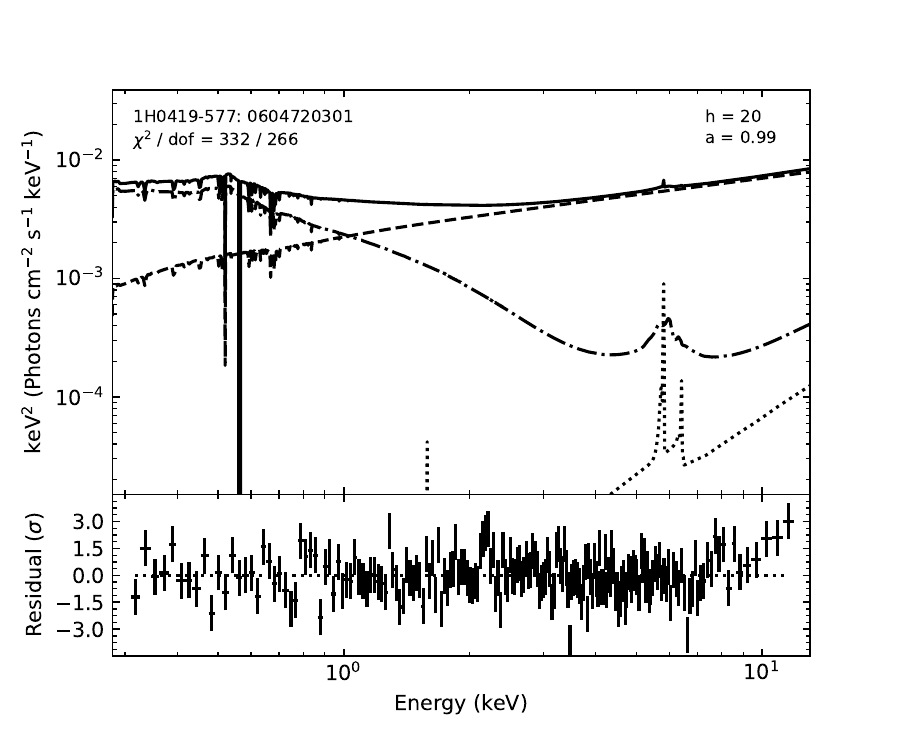}
\includegraphics[width=0.49\textwidth]{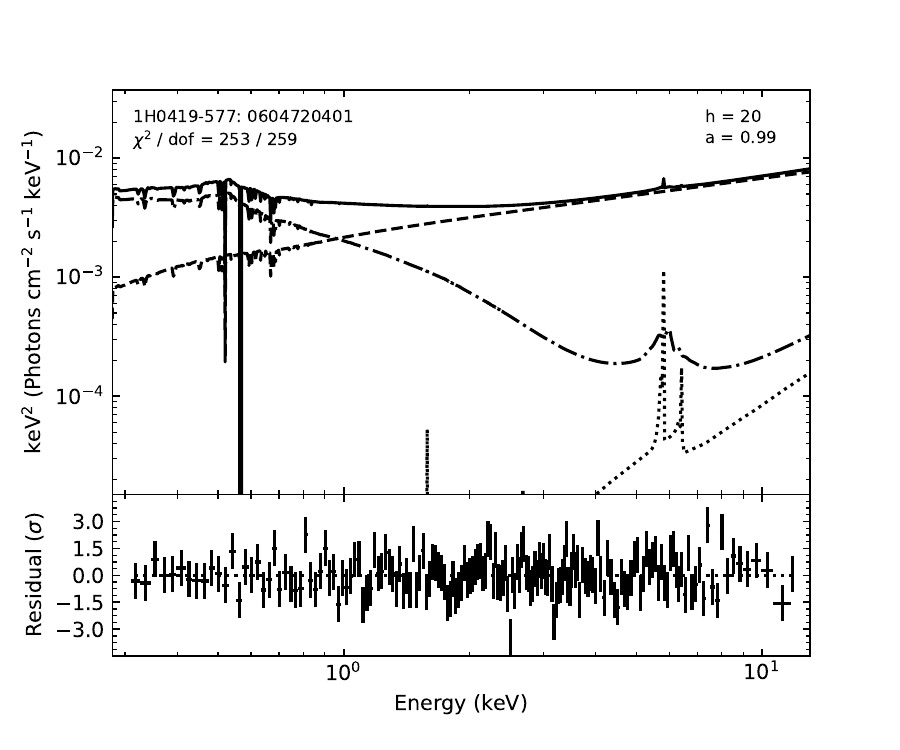}
\includegraphics[width=0.49\textwidth]{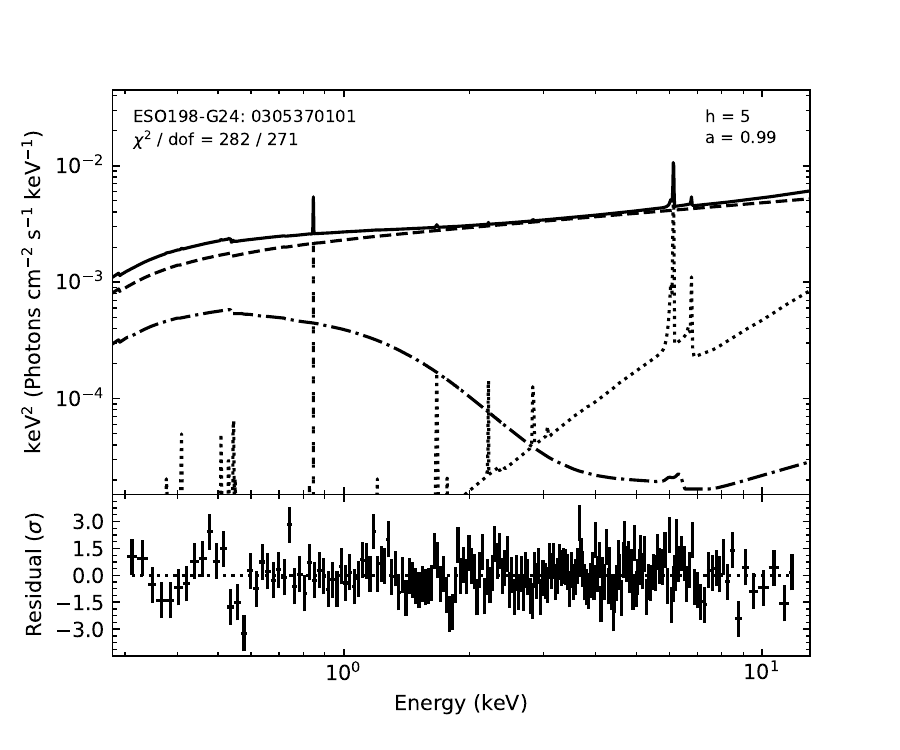}
\includegraphics[width=0.49\textwidth]{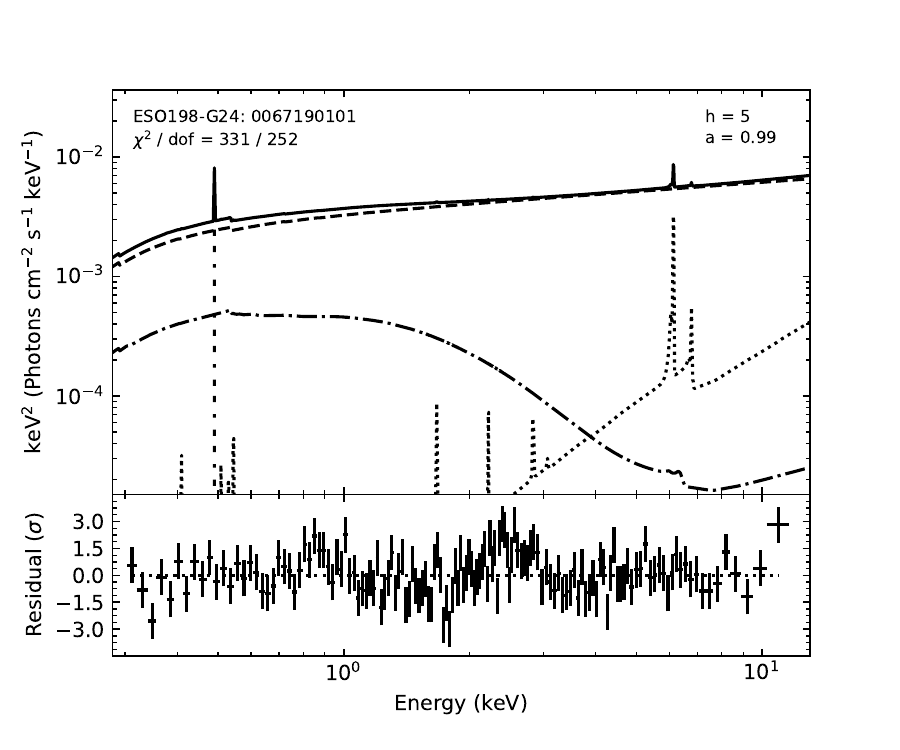}
\includegraphics[width=0.49\textwidth]{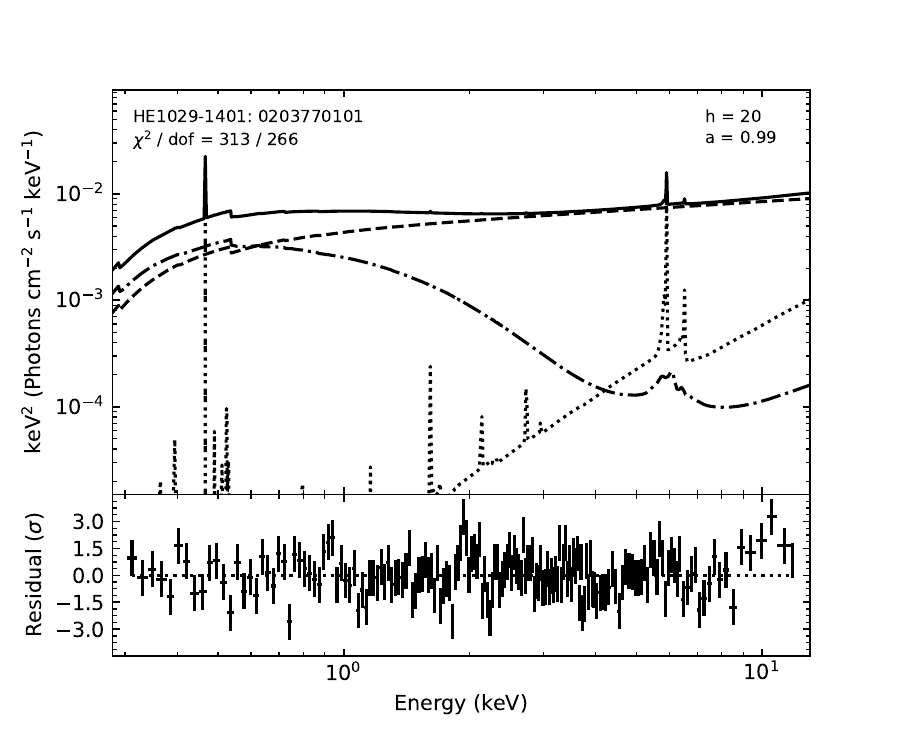}
\includegraphics[width=0.49\textwidth]{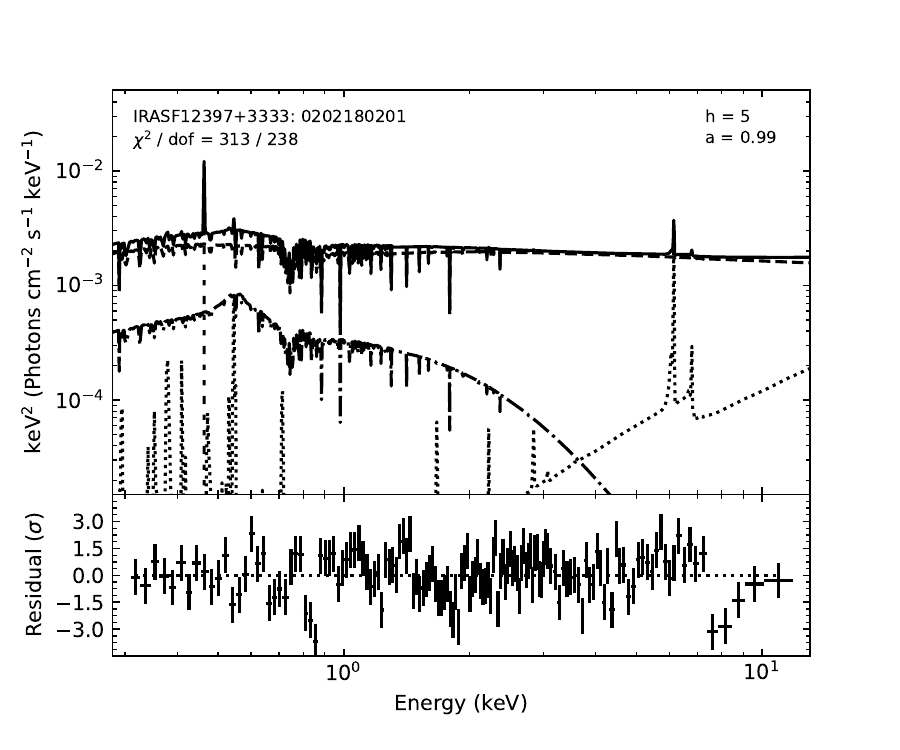}
\caption{The upper part of each panel plots the spectral model found from the best fit to each observation in our sample (Tables~\ref{table:bestfits}--\ref{table:WA}). The solid line shows the total model, while the short-dashed, dot-dashed, and dotted lines plot the \textit{power-law}, \rexcor\ and \textit{xillver} components, respectively. Any Gaussian lines are shown using dot-dot-dashed lines.}
\label{fig:bestfits}
\end{figure*}

\begin{figure*}
\includegraphics[width=0.49\textwidth]{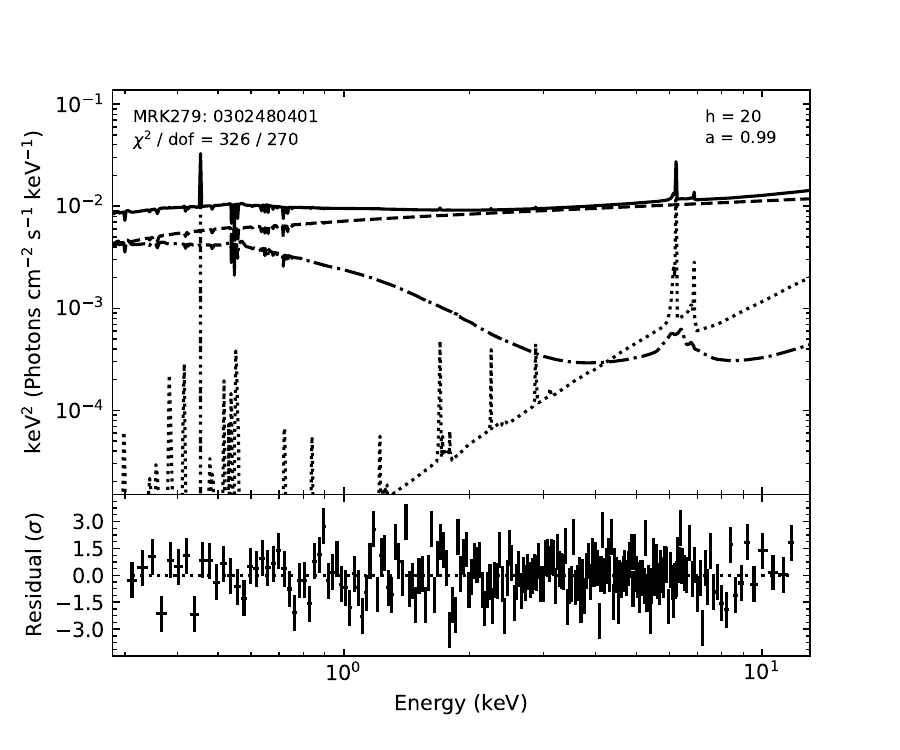}
\includegraphics[width=0.49\textwidth]{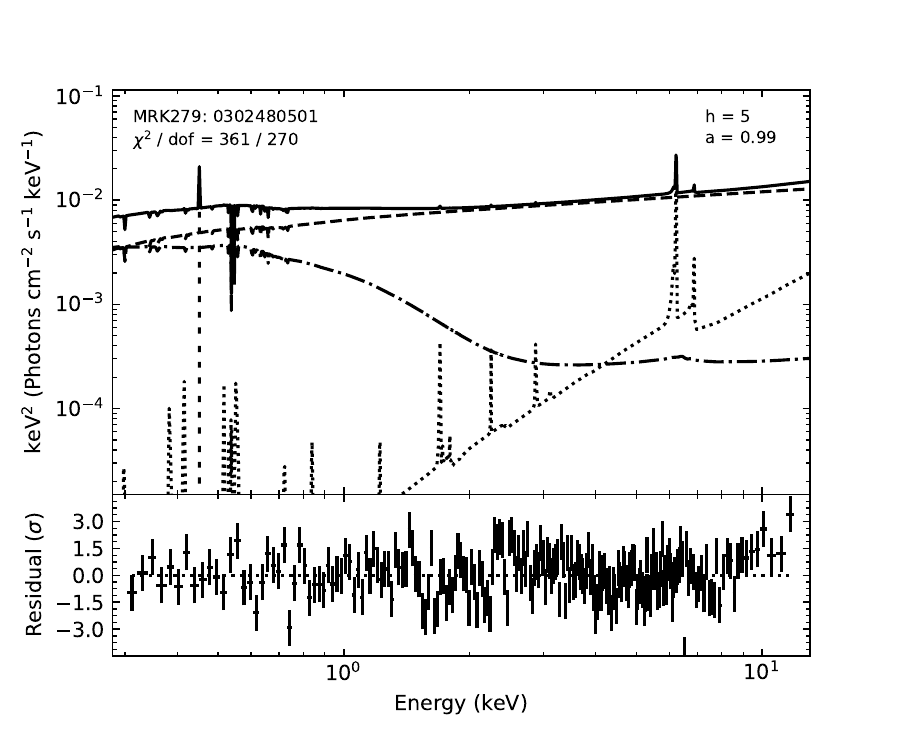}
\includegraphics[width=0.49\textwidth]{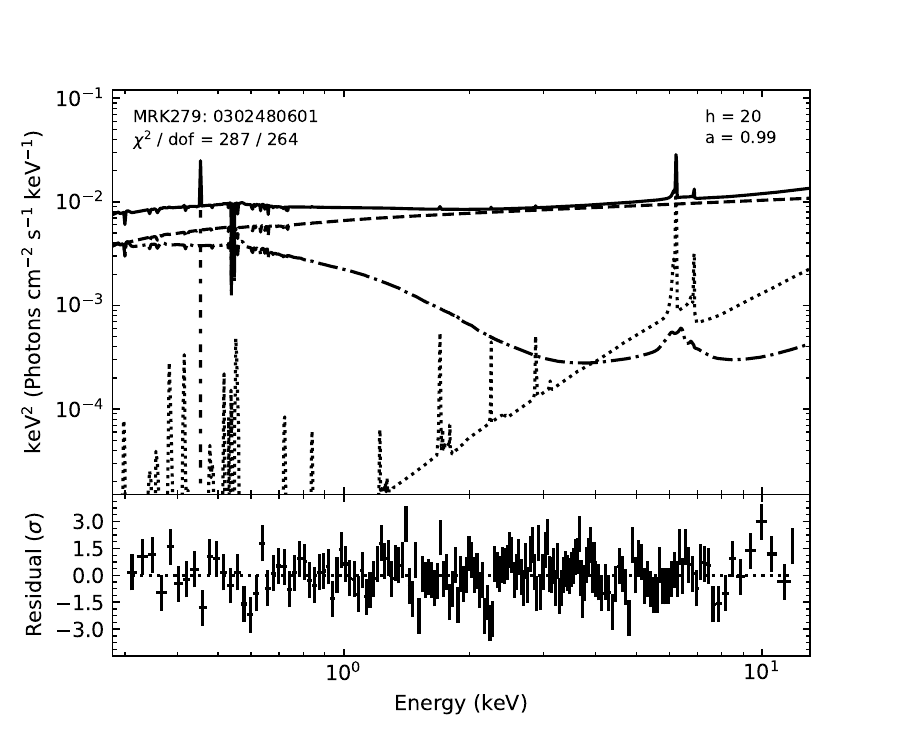}
\includegraphics[width=0.49\textwidth]{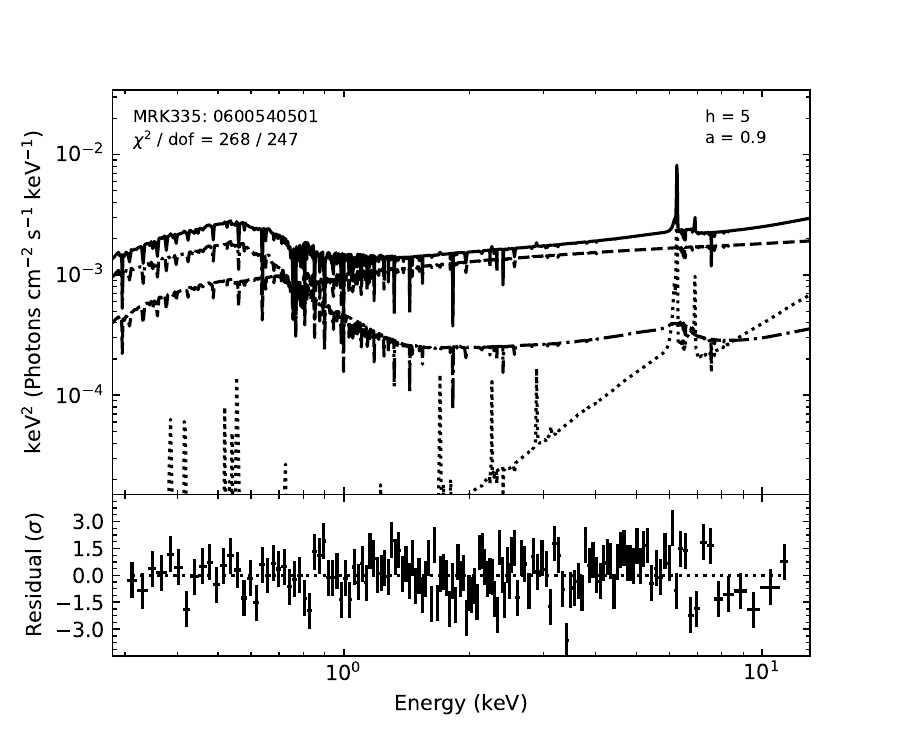}
\includegraphics[width=0.49\textwidth]{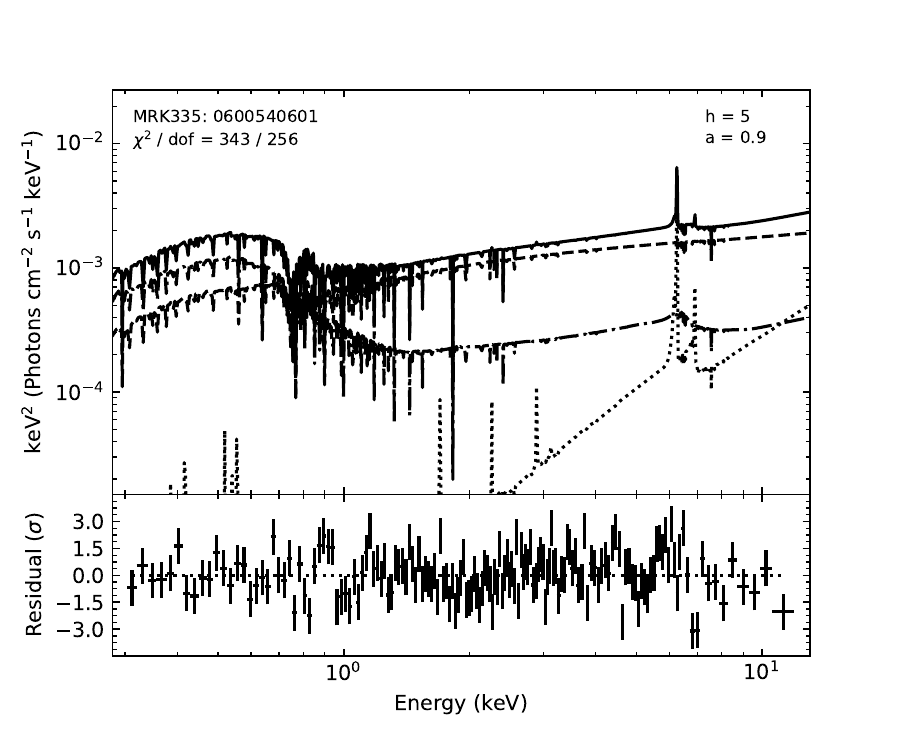}
\includegraphics[width=0.49\textwidth]{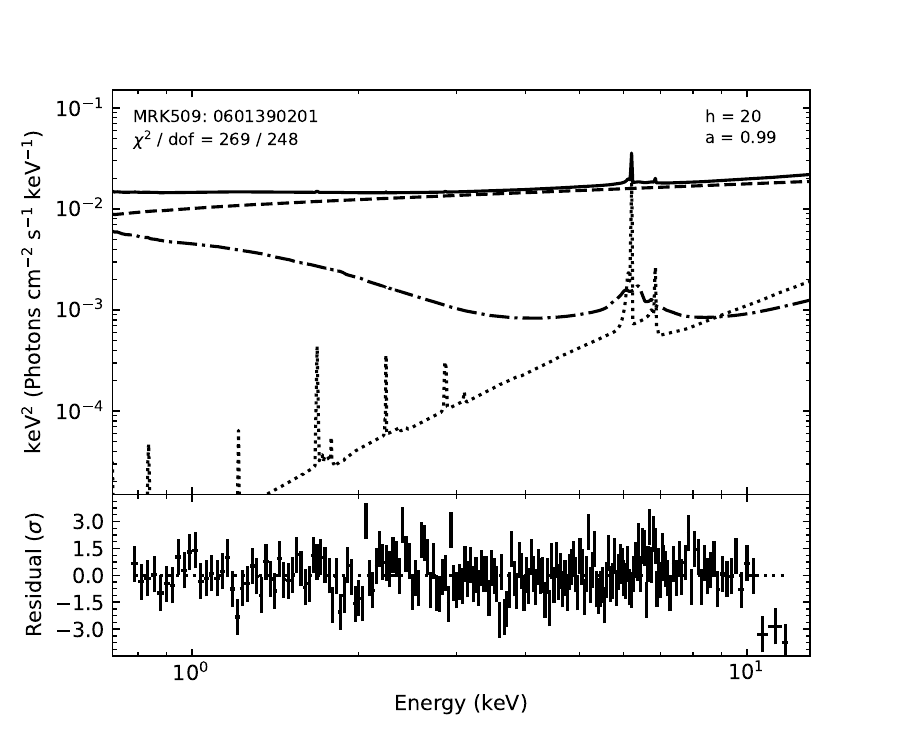}
\contcaption{}
\end{figure*}

\begin{figure*}
\includegraphics[width=0.49\textwidth]{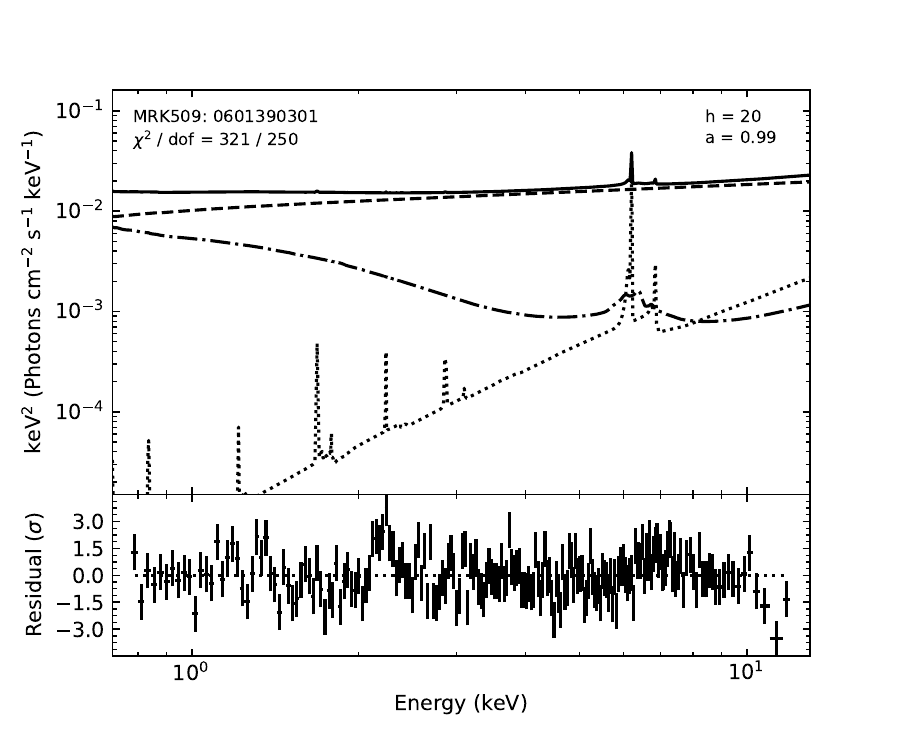}
\includegraphics[width=0.49\textwidth]{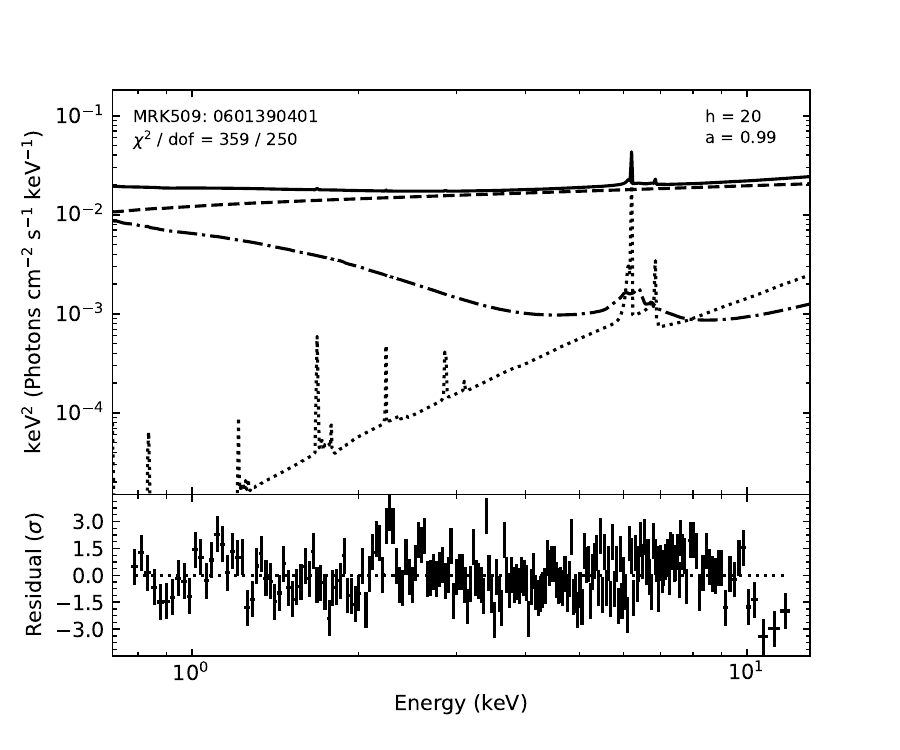}
\includegraphics[width=0.49\textwidth]{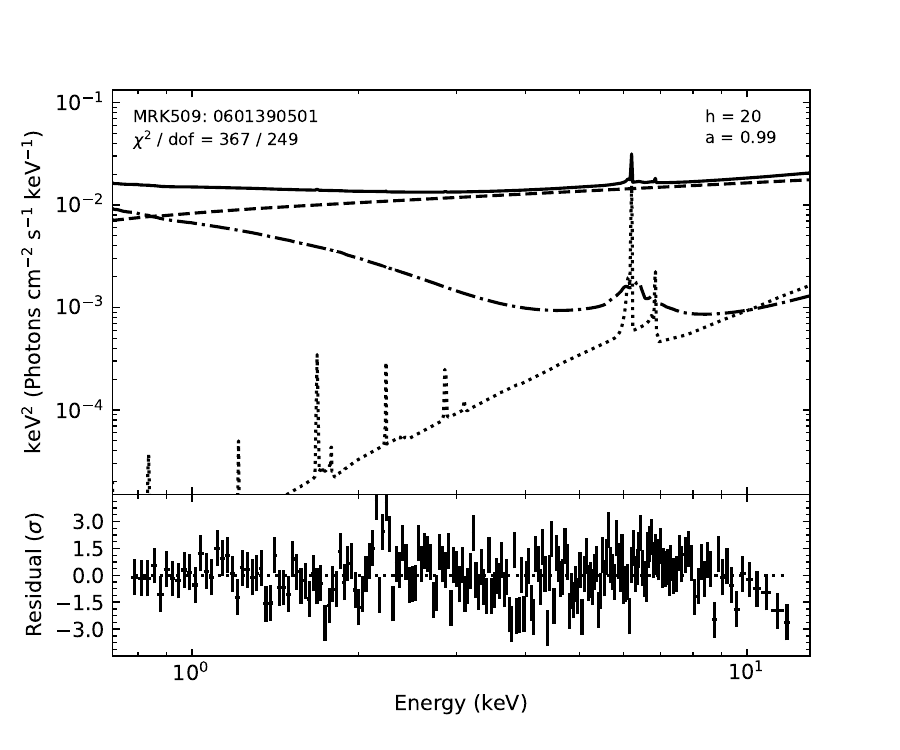}
\includegraphics[width=0.49\textwidth]{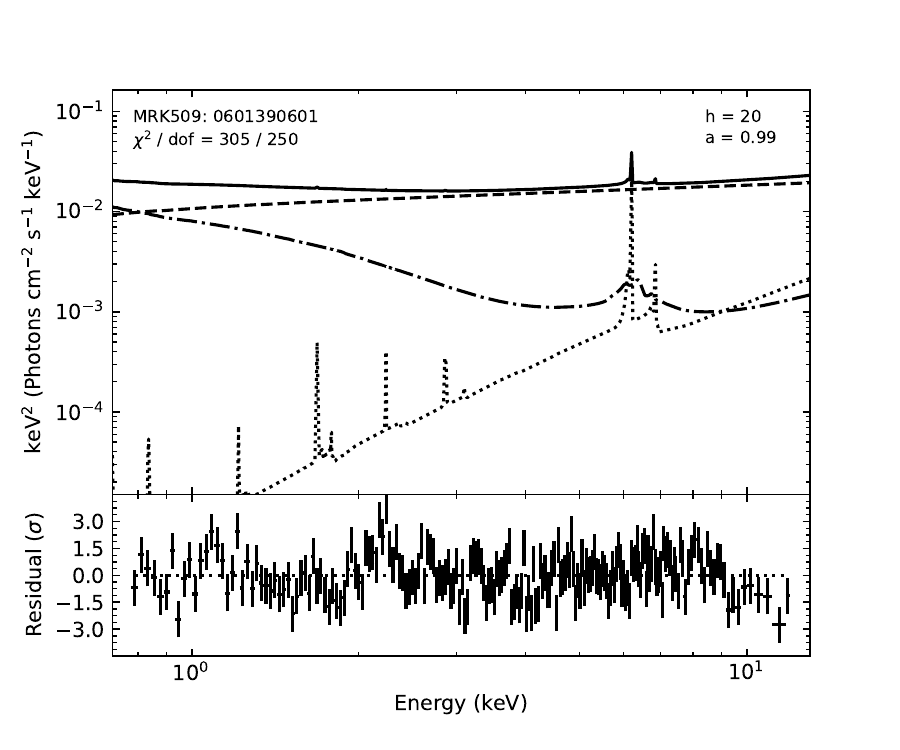}
\includegraphics[width=0.49\textwidth]{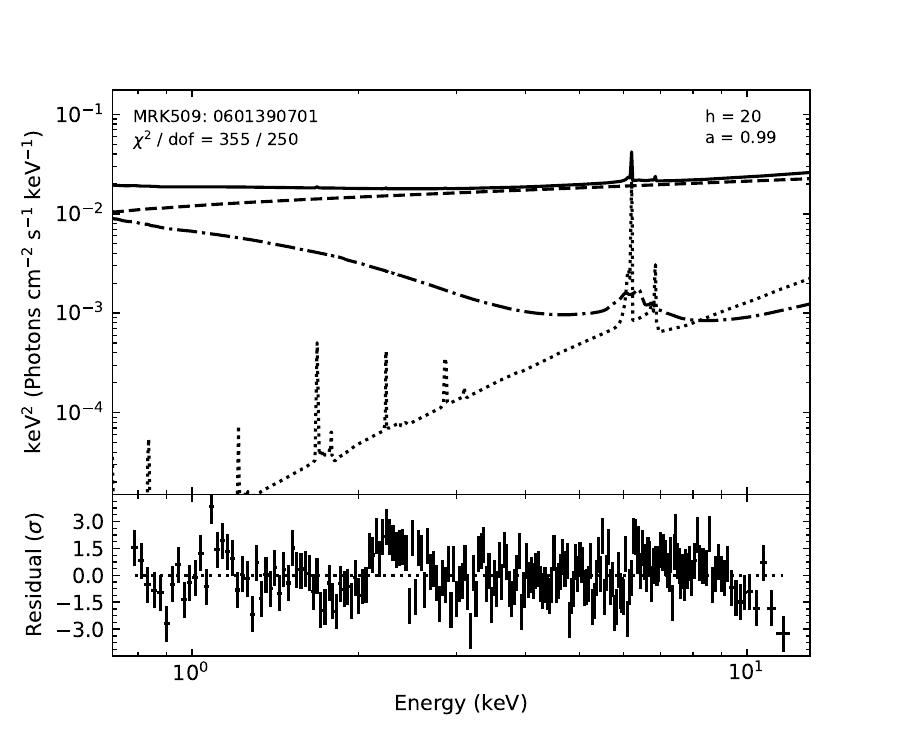}
\includegraphics[width=0.49\textwidth]{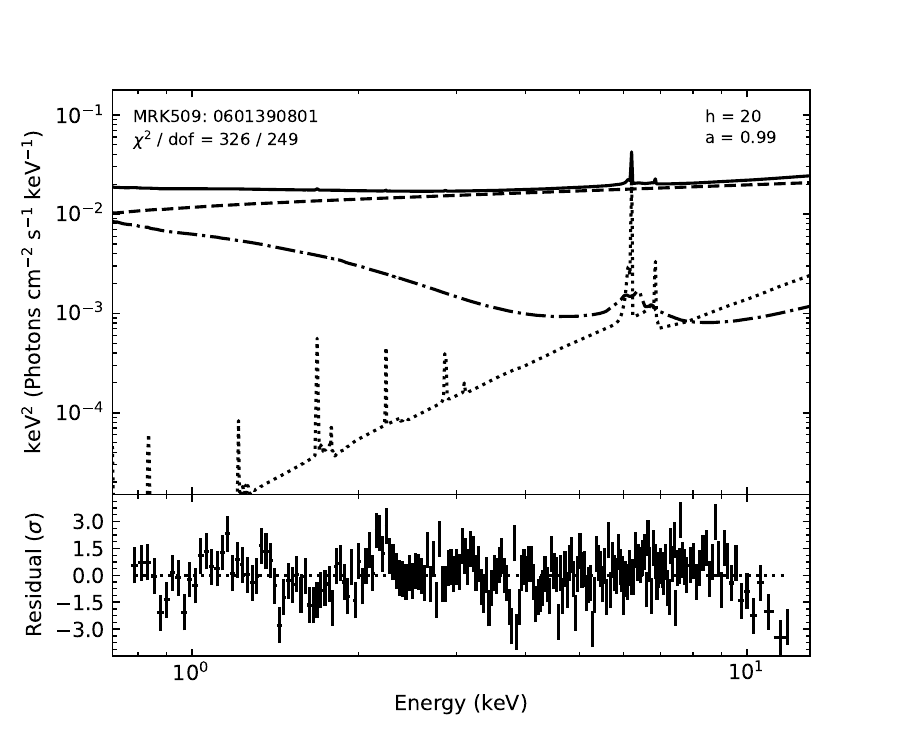}
\contcaption{}
\end{figure*}

\begin{figure*}
\includegraphics[width=0.49\textwidth]{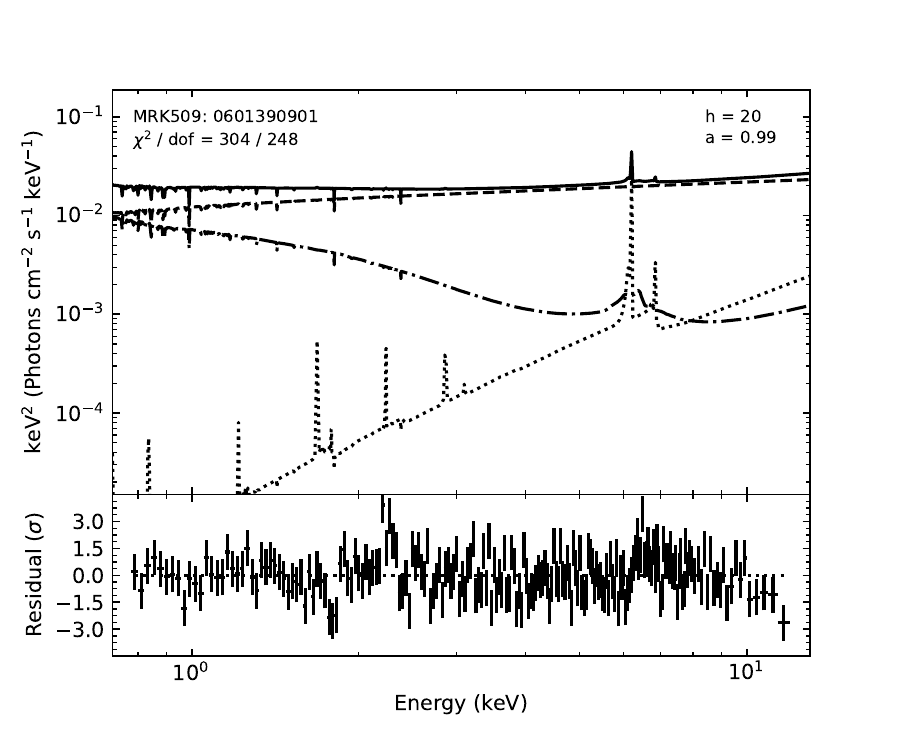}
\includegraphics[width=0.49\textwidth]{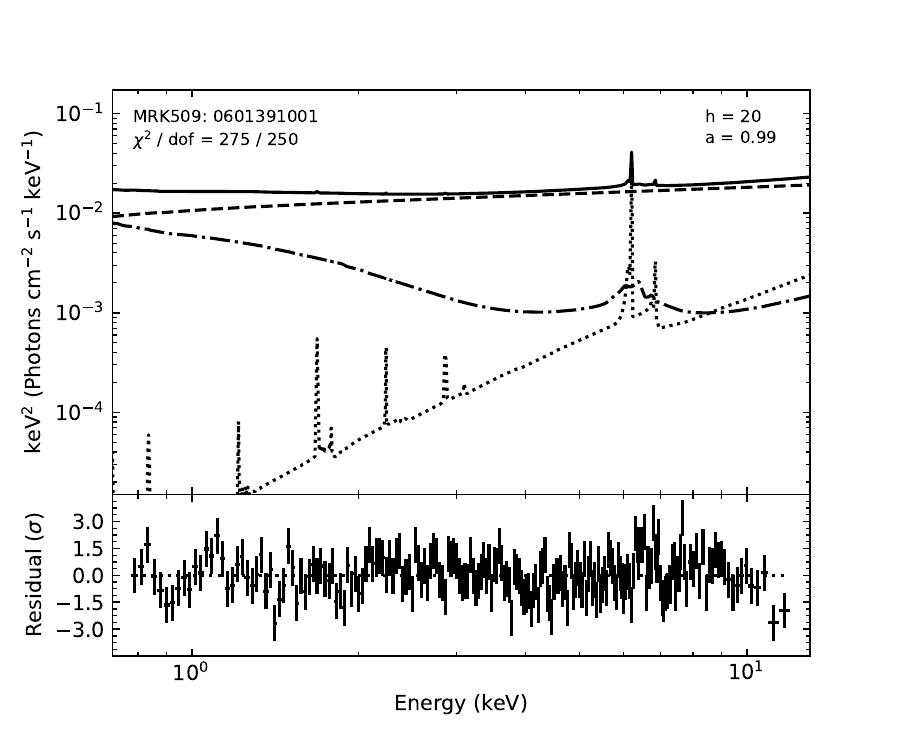}
\includegraphics[width=0.49\textwidth]{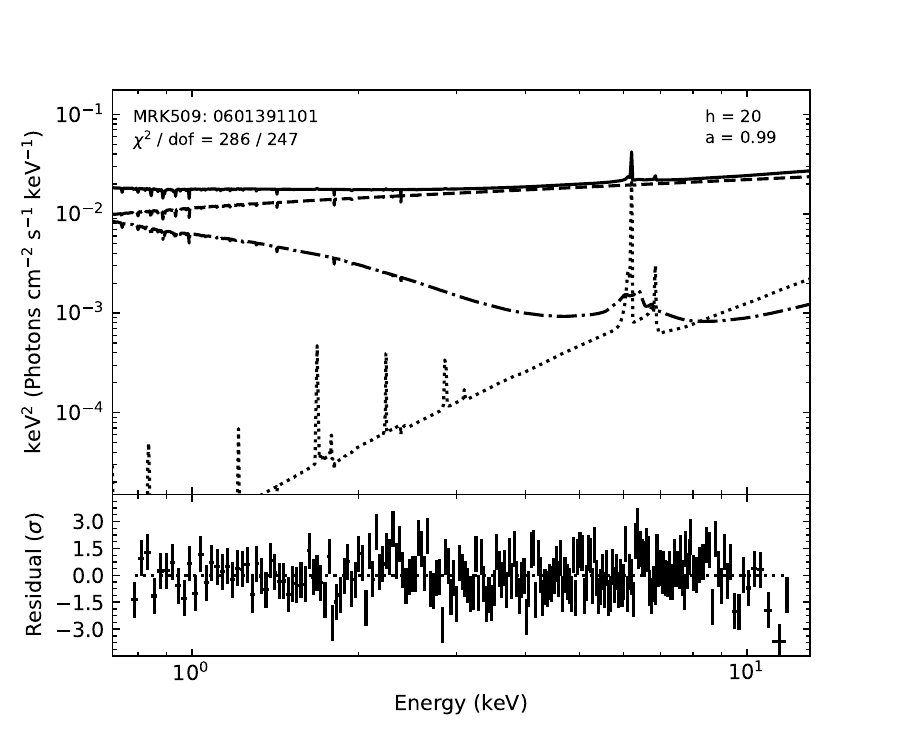}
\includegraphics[width=0.49\textwidth]{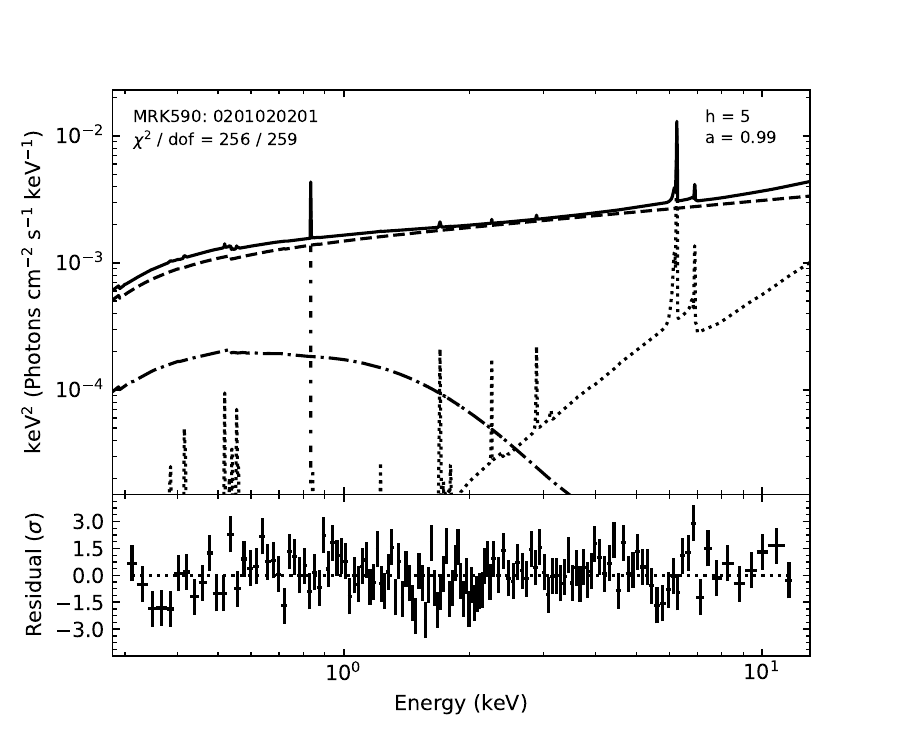}
\includegraphics[width=0.49\textwidth]{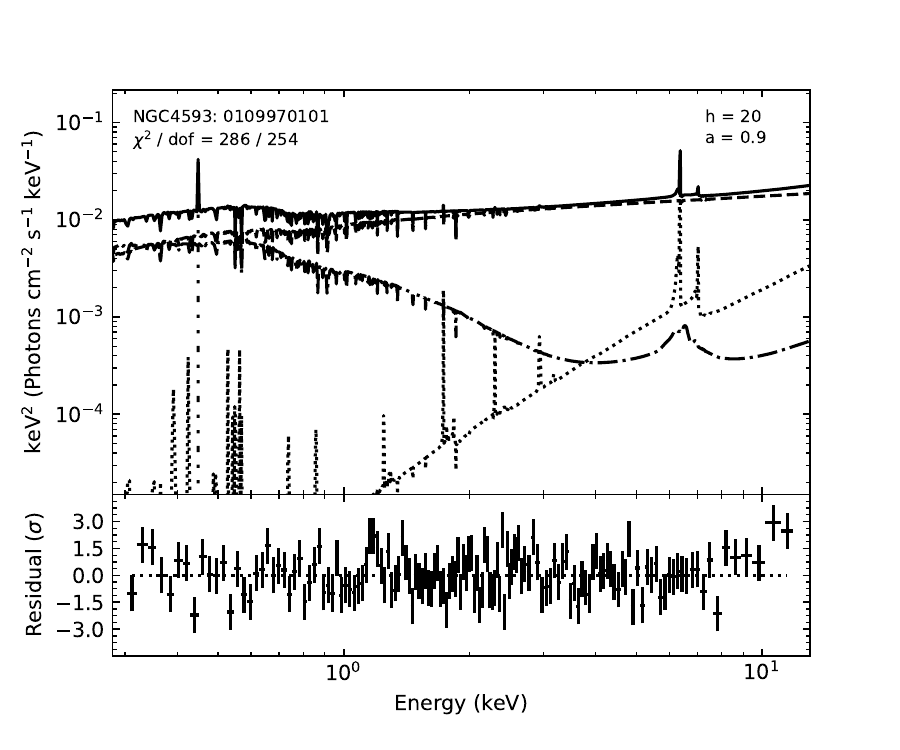}
\includegraphics[width=0.49\textwidth]
                {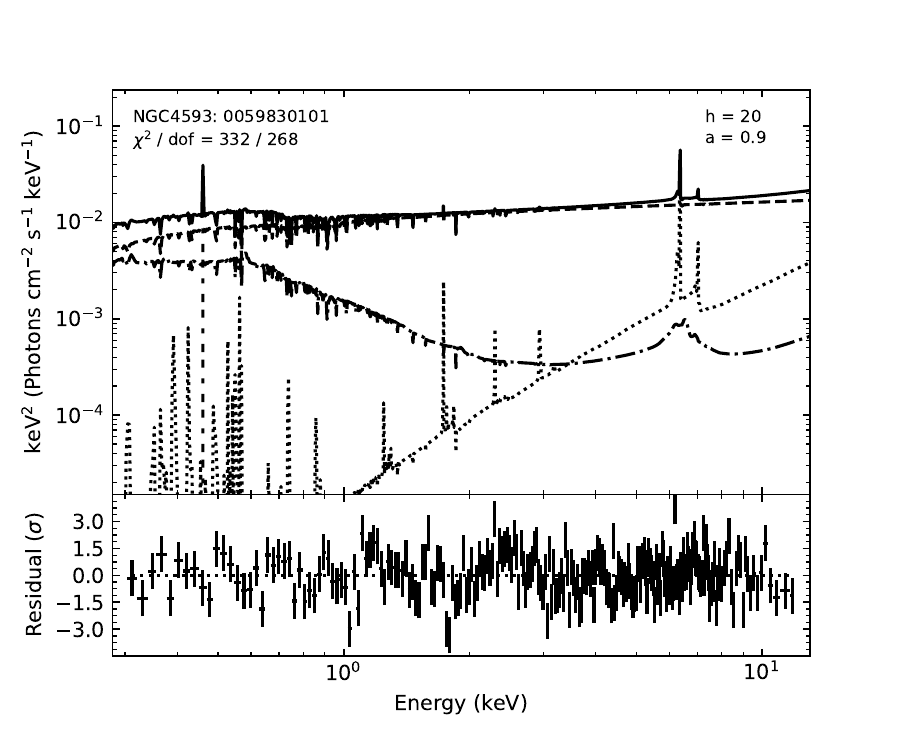}
\contcaption{}
\end{figure*}

\begin{figure*}
\includegraphics[width=0.49\textwidth]{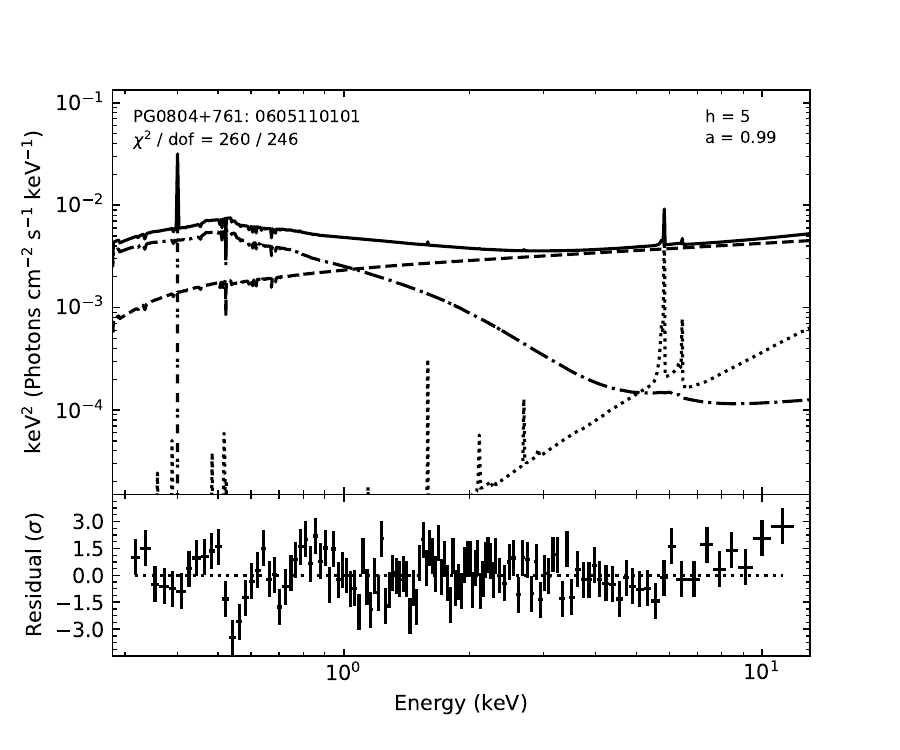}
\includegraphics[width=0.49\textwidth]{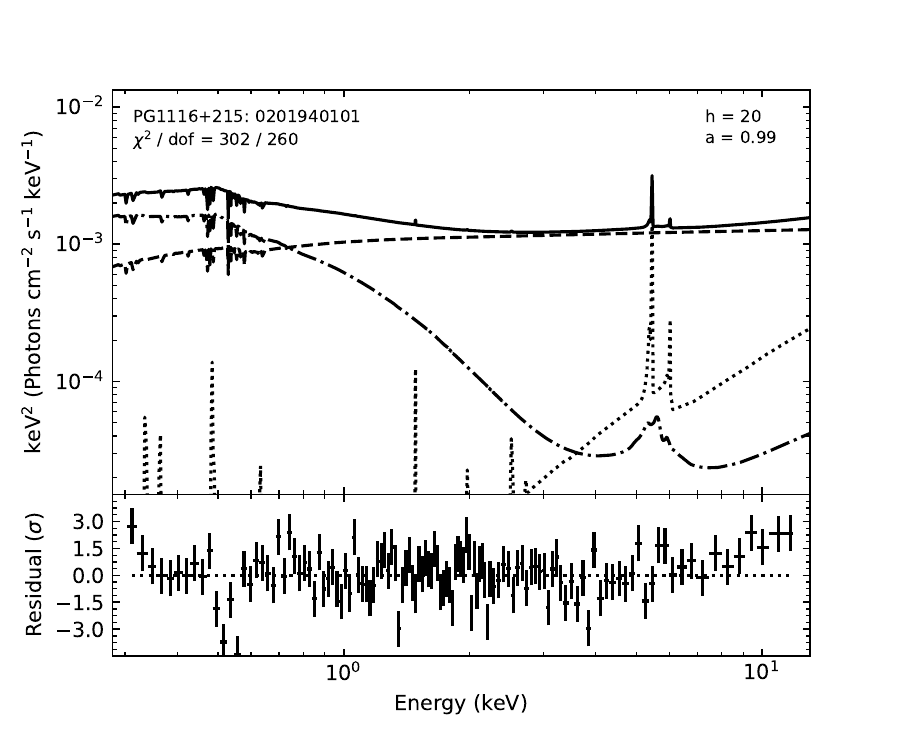}
\includegraphics[width=0.49\textwidth]{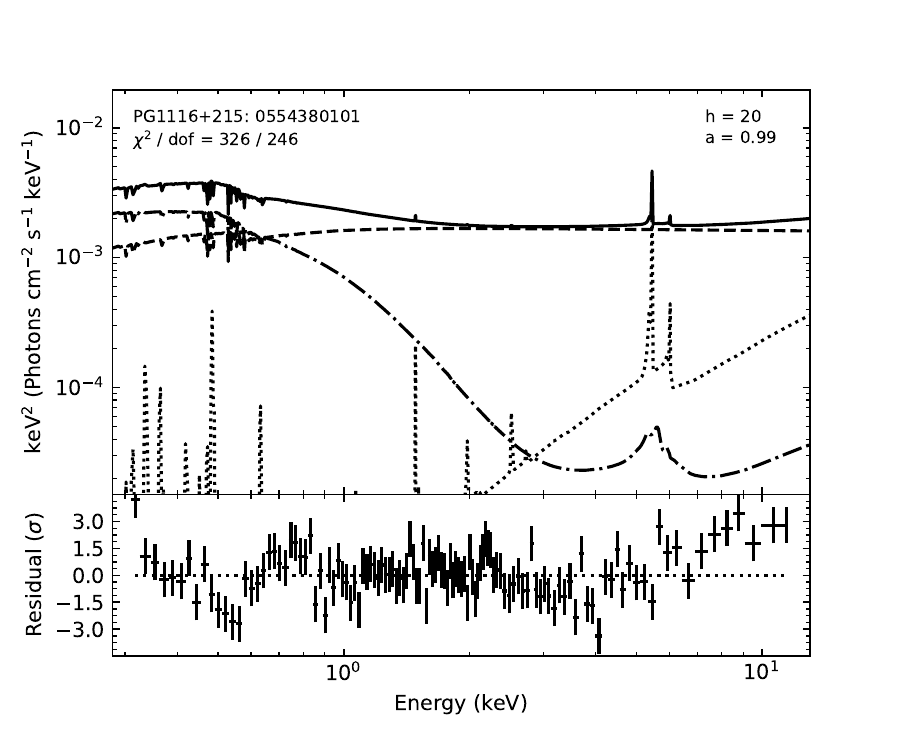}
\includegraphics[width=0.49\textwidth]{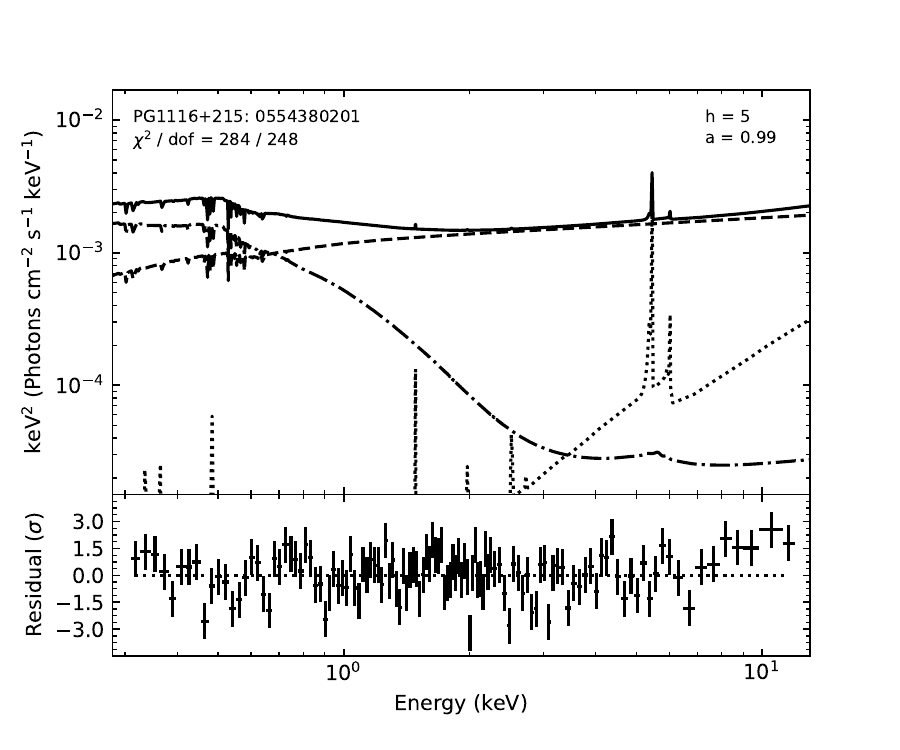}
\includegraphics[width=0.49\textwidth]{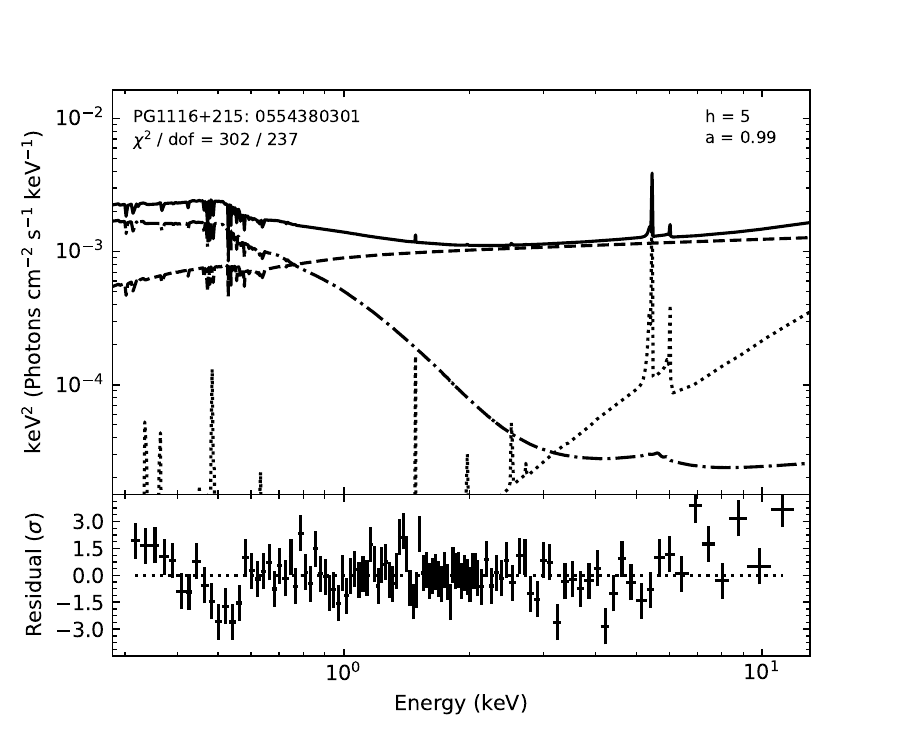}
\includegraphics[width=0.49\textwidth]{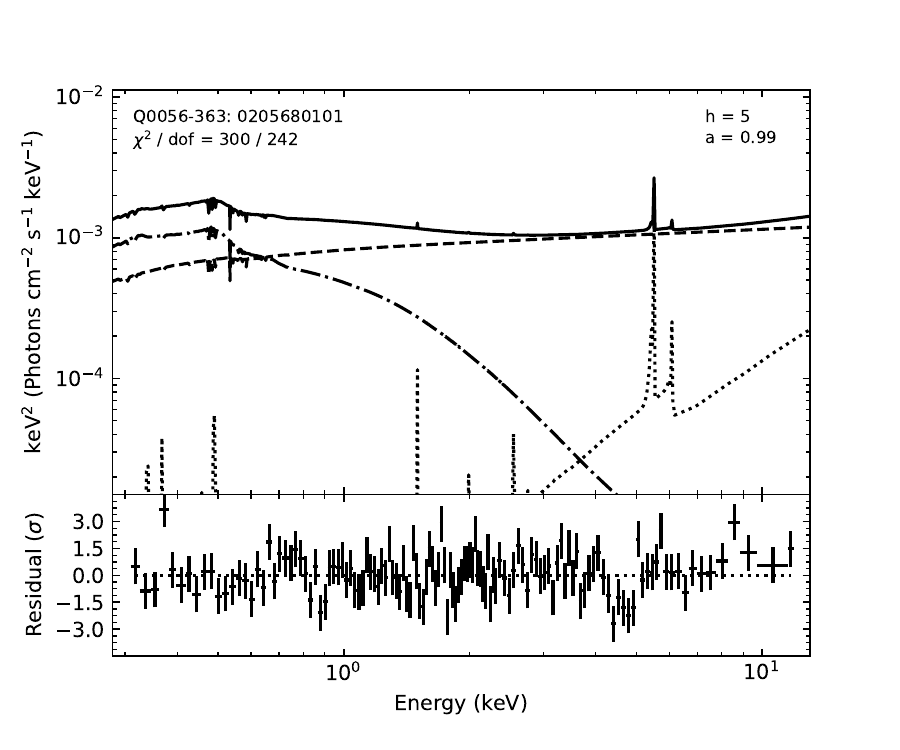}
\contcaption{}
\end{figure*}

\begin{figure*}
\includegraphics[width=0.49\textwidth]{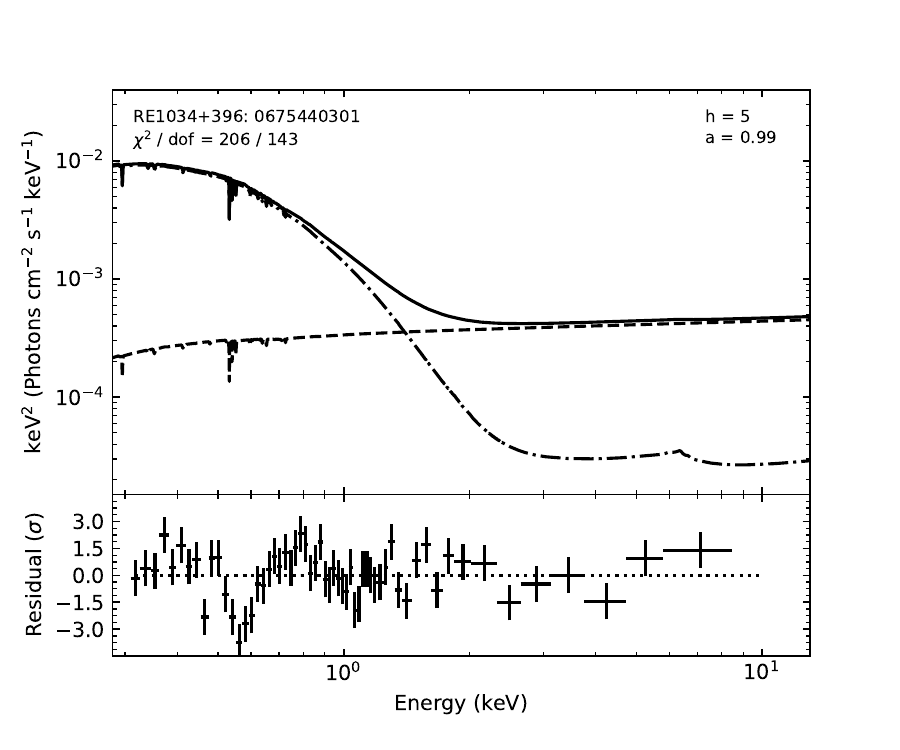}
\includegraphics[width=0.49\textwidth]{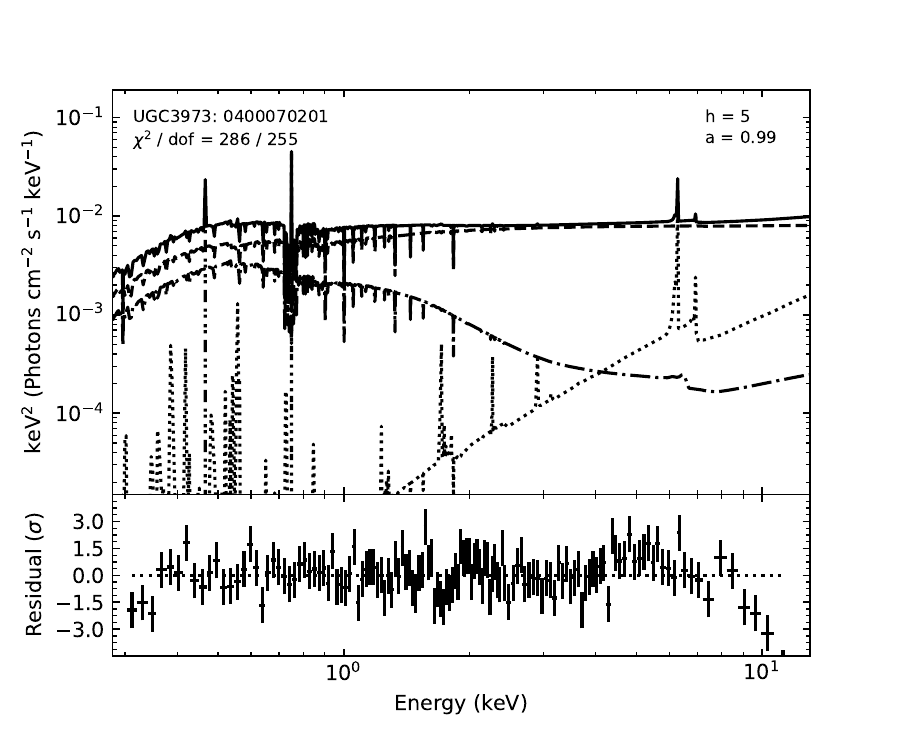}
\includegraphics[width=0.49\textwidth]{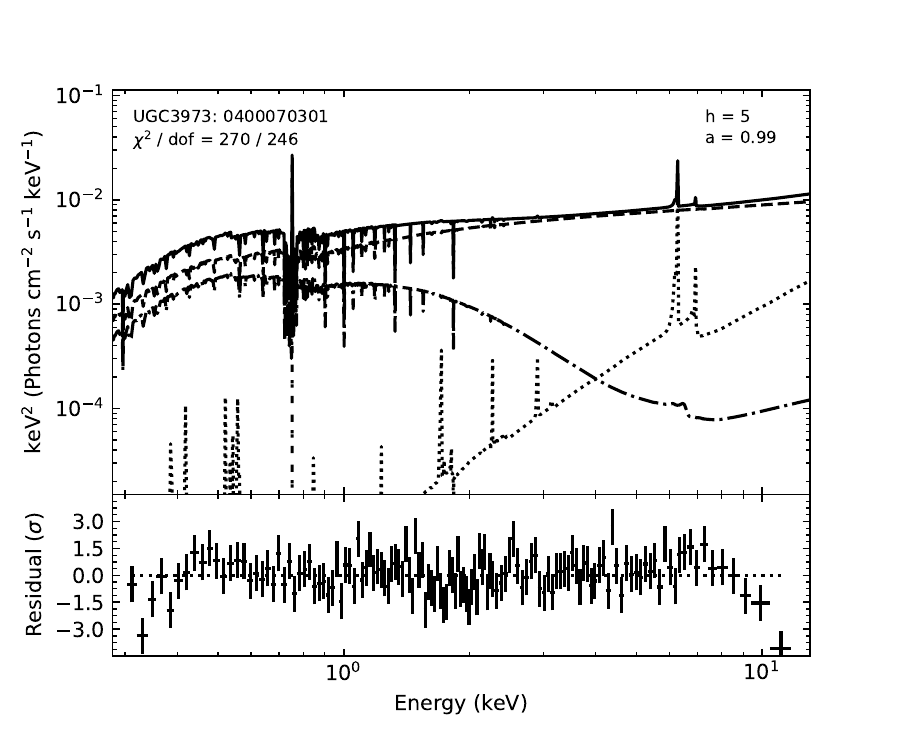}
\includegraphics[width=0.49\textwidth]{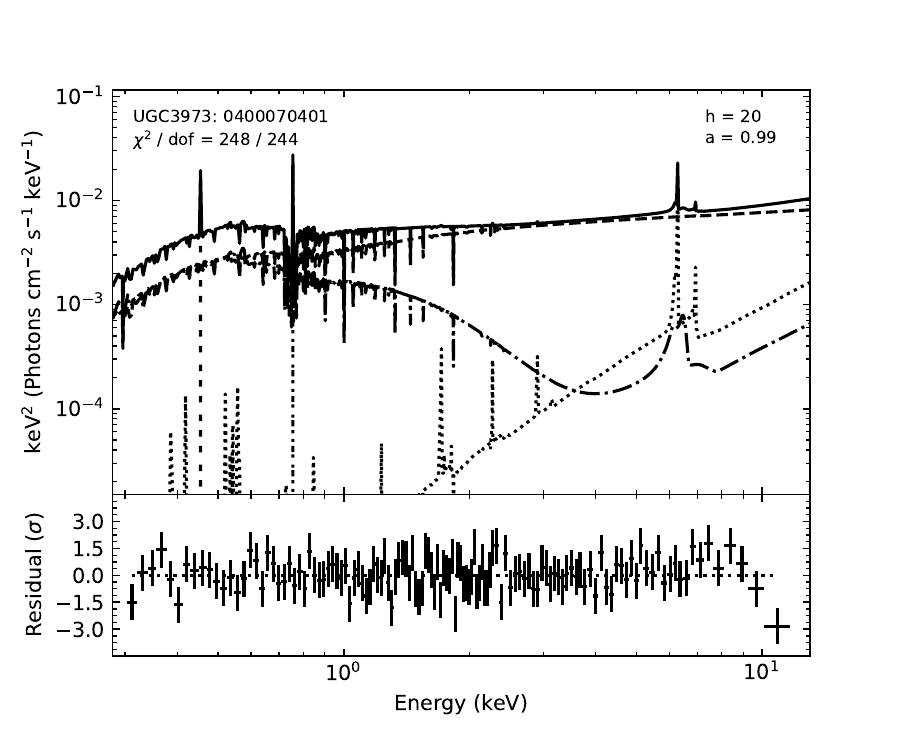}
\contcaption{}
\end{figure*}



\section{Model Dependent Effects}
\label{app:C}
The results presented in Section~\ref{sect:res} focus on the \rexcor\ parameters related to the distribution of accretion energy. However, the \rexcor\ models also depend on the assumed Eddington ratio, the height of the hot corona (situated as a `lamppost' above the black hole) and the spin of the black hole. In this Appendix, we discuss how our results may depend on these other parameters.

\subsection{The Eddington ratio, $\lambda$}
\label{sub:lambda}
As mentioned in Sect.~\ref{sect:sample}, we used the $\lambda=0.01$ \rexcor\ grids when fitting observations with $\lambda_{\mathrm{obs}} < 0.05$. Following the predictions of standard accretion theory \citep[e.g.,][]{ss73}, the $\lambda=0.01$ \rexcor\ models are calculated with a denser accretion disc than the $\lambda=0.1$ models. This fact, combined with the lower luminosity, predicts a less ionized reflection and emission spectrum for a given $f_X$ and $h_f$ \citep{xb22}. Therefore, it is possible that the high values of $h_f$ and $f_X$ found in observations with $\lambda_{\mathrm{obs}} < 0.05$ are just a result of this choice and are not required by the data. 

To determine how sensitive the warm corona parameters are to the assumed Eddington ratio of the \rexcor\ models, we re-fit all 7 observations which were originally fit with the $\lambda=0.01$ models with the lower density $\lambda=0.1$ \rexcor\ grids, and the results are tabulated in Tables~\ref{table:bf2}--\ref{table:WA2} which follow the same format as Tables~\ref{table:bestfits}--\ref{table:WA}. The right-most column of Table~\ref{table:bf2} shows the change in $\chi^2$, as well as any difference in the degrees-of-freedom, when the $\lambda=0.01$ \rexcor\ grids are replaced with $\lambda=0.1$ grids. To determine how significant these different fits are from the original ones, we use the Bayesian Information Criterion (BIC; \citealt{schwartz78,liddle04}), which for $\chi^2$ statistics is $\mathrm{BIC}=\chi^2 + k \ln(N)$, where $k$ is the number of parameters and $N$ is the number of data points \citep[e.g.,][]{yamada23}. If the $\Delta \mathrm{BIC}$ between two models is $ < 6$ then there is only marginal evidence for one model being preferred over the other \citep{mukherjee98}. In all of our cases, $N$ is the same between the two models, and only two observations (MRK590 and UGC3973 [Obs. ID 0400070301]) have a different number of parameters. For the remaining 5 observations the $\Delta \mathrm{BIC}$ is equivalent to $\Delta \chi^2$.

\begin{table*}
    \centering
    \caption{As in Table~\ref{table:bestfits}, but now showing the results when the $\lambda=0.1$ \rexcor\ grids are used when fitting the 7 observations of AGNs that were originally fit with $\lambda=0.01$ \rexcor\ models. The last column shows the change in $\chi^2$ and degrees of freedom that result from using the $\lambda=0.1$ \rexcor\ grids.}    
    \label{table:bf2}
    \begin{tabular}{l|l|c|c|c|c|c|c|c}
 Object & Obs.\ ID & $a$ & $h$ ($r_g$) & $h_f$ & $f_X$ & $\tau$ & $\Gamma$ & $\Delta \chi^2/\Delta \mathrm{dof}$ \\ \hline
ESO198-G24 & 0305370101 & $0.99$ & $5$ & $0.13_{-0.04}^{+0.12}$ & $0.15_{-0.03}^{+0.05p}$ & $10.5_{-0.5p}^{+1.8}$ & $1.72_{-0.02}^{+0.02}$ & $-8/0$ \\ \vspace{2mm}
 & 0067190101 & & $5$ & $0.72_{-0.22}^{+0.06}$ & $0.20_{-0.09}^{+0p}$ & $30.0_{-13.7}^{+0p}$ & $1.76_{-0.03}^{+0.04}$ & $-7/0$ \\ \vspace{2mm}
MRK590 & 0201020201 & $0.99$ & $5$ & $0.02^{+0.06}_{-0.02p}$ & $0.20_{-0.04}^{+0p}$ & $10.0_{-0p}^{+0.5}$ & $1.69_{-0.03}^{+0.02}$ & $-9/+2$\\ \vspace{2mm}
Q0056-363 & 0205680101 & $0.99$ & $20$ & $0.55\pm 0.01$ & $0.020^{+0.008}_{-0p}$ & $10.5^{+0.4}_{-0.5p}$ & $1.77^{+0.08}_{-0.06}$ & $+6/0$ \\ 
UGC3973	& 0400070201 & $0.99$ & $20$ & $0.65^{+0.03}_{-0.09}$ & $0.20_{-0.01}^{+0p}$ & $27.6^{+0.5}_{-2.2}$ & $2.03^{+0.01}_{-0.02}$ & $-5/0$\\
    & 0400070301 &  & $20$ & $0.71\pm 0.01$ & $0.20_{-0.05}^{+0p}$ & $30.0_{-2.7}^{+0p}$ & $1.77^{+0.04}_{-0.03}$ & $-17/-2$\\
    & 0400070401 &  & $20$ & $0.32_{-0.09}^{+0.26}$ & $0.19_{-0.04}^{+0.01p}$ & $13.4_{-0.3}^{+4.8}$ & $1.81_{-0.06}^{+0.03}$ & $+8/0$\\
 \end{tabular}
 \end{table*}

\begin{table*}
    \centering
    \caption{As in Table~\ref{table:fluxes}, but now showing the results when the $\lambda=0.1$ \rexcor\ grids are used when fitting the 7 observations that were originally fit with the $\lambda=0.01$ grids. Note that MRK590 no longer requires a Gaussian emission line.}  
    \label{table:fluxes2}
    \begin{tabular}{l|l|c|c|c|c|c|c|c}
 Object & Obs.\ ID & $\log F$ (\rexcor) & $\log F$ (\textit{power-law}) & $\log F$ (\textit{xillver}) & $E$ (keV) & $K$ ($\times 10^{-5}$) & $\sigma$ (keV) \\ 
 \hline
ESO198-G24 & 0305370101 & $-11.82\pm 0.07$ & $-10.78\pm 0.01$ & $-12.26^{+0.07}_{-0.08}$ & $0.50\pm 0.01$ & $5.1^{+1.8}_{-1.7}$ & $0^f$ \\ \vspace{2mm}
 & 0067190101 & $ -11.91_{-0.19}^{+0.12}$ & $-10.64\pm 0.01$ & $-12.57_{-0.39}^{+0.20}$ & $0.51\pm 0.01$ & $9.9^{+3.0}_{-3.2}$ & $0^f$\\ \vspace{2mm}
MRK590 & 0201020201 & $-12.15_{-0.10}^{+0.11}$ & $-10.97\pm 0.01$ & $-12.24_{-0.08}^{+0.06}$ & -- & -- & -- \\ \vspace{2mm}
 Q0056-363 & 0205680101 & $-11.49^{+0.04}_{-0.06}$ & $-11.33_{-0.02}^{+0.03}$ & $-12.93^{+0.15}_{-0.18}$ & -- & -- & --\\ 
UGC3973	& 0400070201 & $-10.97^{+0.01}_{-0.03}$ & $-10.34\pm 0.01$ & $-11.83_{-0.11}^{+0.06}$ & $0.48\pm 0.01$ & $52^{15}_{-8}$ & $0^f$\\
& & & & & $0.77\pm 0.01$ & $229^{+54}_{-83}$ & $0^f$\\
    & 0400070301 & $-11.13_{-0.04}^{+0.07}$ & $-10.47\pm 0.01$ & $-11.97_{-0.14}^{+0.10}$ & $0.50\pm 0.02$ & $18^{+12}_{-7}$ & $0^f$ \\
 & & & & & $0.77_{-0.02}^{+0.01}$ & $132^{+63}_{-16}$ & $0^f$\\
    & 0400070401 & $-10.91^{+0.04}_{-0.02}$ & $-10.54^{+0.01}_{-0.02}$ & $-11.94_{-0.13}^{+0.10}$ & $0.48\pm 0.01$ & $43^{+12}_{-9}$ & $0^f$\\
& & & & & $0.77\pm 0.01$ & $36^{+20}_{-6}$ & $0^f$\\

 \end{tabular}
 \end{table*}

\begin{table}
    \centering
    \caption{As in Table~\ref{table:WA}, but now showing the results when the $\lambda=0.1$ \rexcor\ grids are used when fitting the 7 observations that were originally fit with the $\lambda=0.01$ models. }  
    \label{table:WA2}
    \begin{tabular}{l|l|c|c}
 Object & Obs.\ ID & $N_H$ & $\log \xi$ \\ 
 \hline
Q0056-363 & 0205680101 & $3.7^{+0.4}_{-0.6}\times 10^{20}$ & $0^{+0.16}_{-0p}$ \\
UGC3973	& 0400070201 & $6.4_{-0.6}^{+0.1} \times 10^{21}$ & $1.69\pm 0.04$\\
    & 0400070301 & $6.5^{+0.3}_{-0.2} \times 10^{21}$ & $1.66_{-0.12}^{+0.02}$\\
    & 0400070401 & $7.6_{-0.9}^{+0.5}\times 10^{21}$ & $1.72^{+0.05}_{-0.03}$\\
 \end{tabular}
 \end{table}

Examination of Table~\ref{table:bf2} shows that using the $\lambda=0.1$ \rexcor\ grids leads to a larger $\chi^2$ for Q0056-363 and one observation of UGC3973 (Obs. ID 0400070401) compared to the fits using the original $\lambda=0.01$ grids. According to the $\Delta {\mathrm{BIC}}$ criterion, the differences in $\chi^2$ are large enough to conclude that the model in Table~\ref{table:bestfits} is preferred. The remaining 5 observations all result in a decrease in $\chi^2$ using the $\lambda=0.1$ grid. Figure~\ref{fig:appC} re-plots the relationships with \lobs\ shown in Sect.~\ref{sect:res} with the \rexcor\ parameters from these five observations replaced with the values from Table~\ref{table:bf2}. 
\begin{figure*}
\centering
    \includegraphics[width=0.475\textwidth]{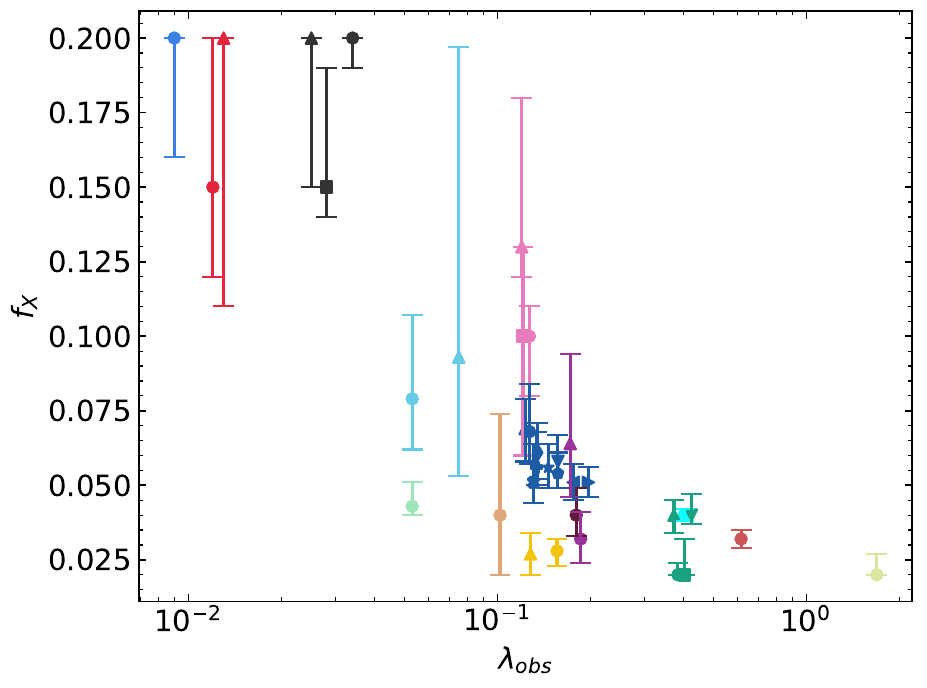}
    \includegraphics[width=0.49\textwidth]{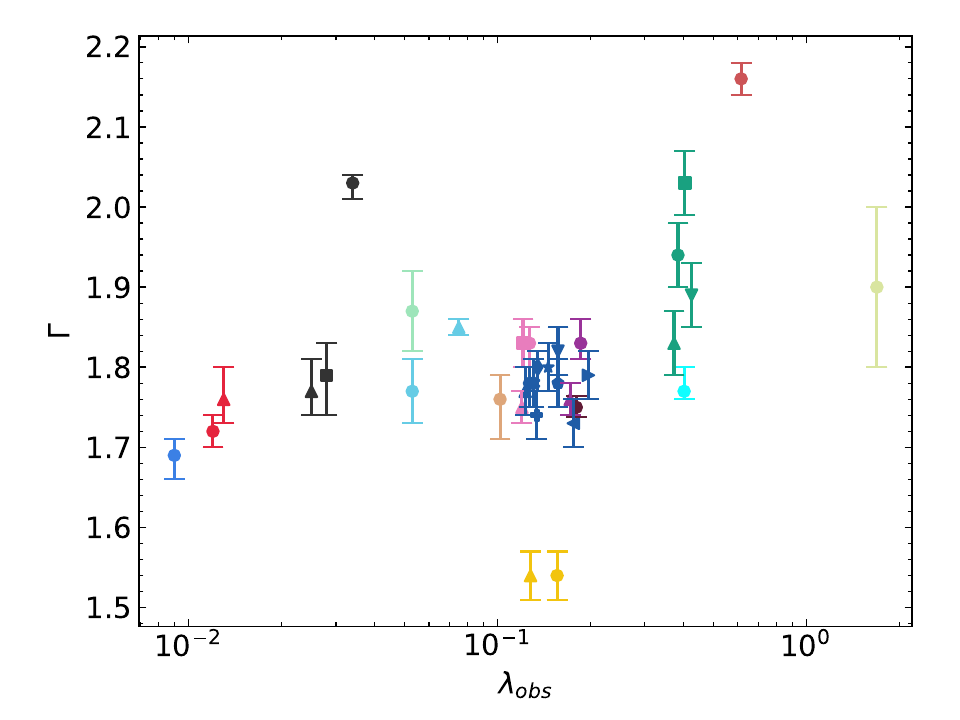}
    \includegraphics[width=0.49\textwidth]
    {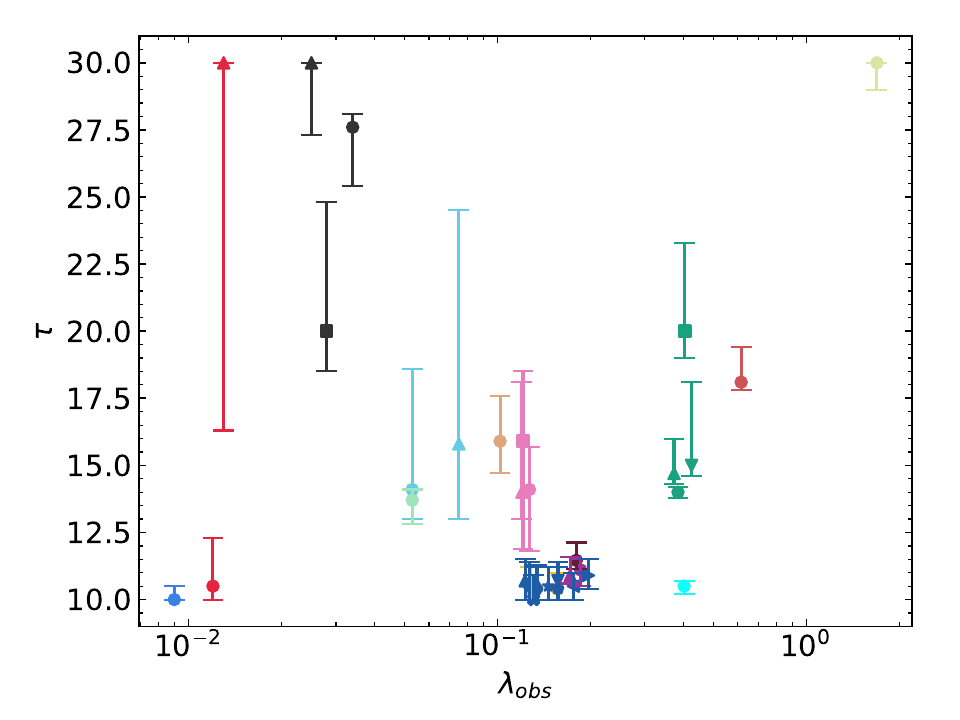}
    \includegraphics[width=0.49\textwidth]
    {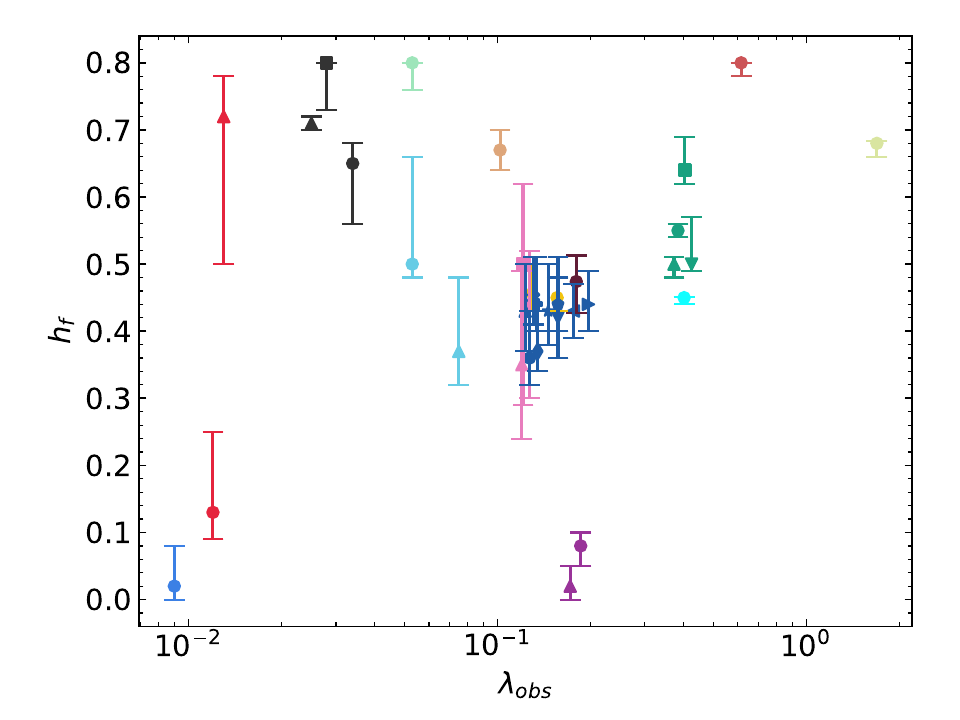}
    \includegraphics[width=0.49\textwidth]{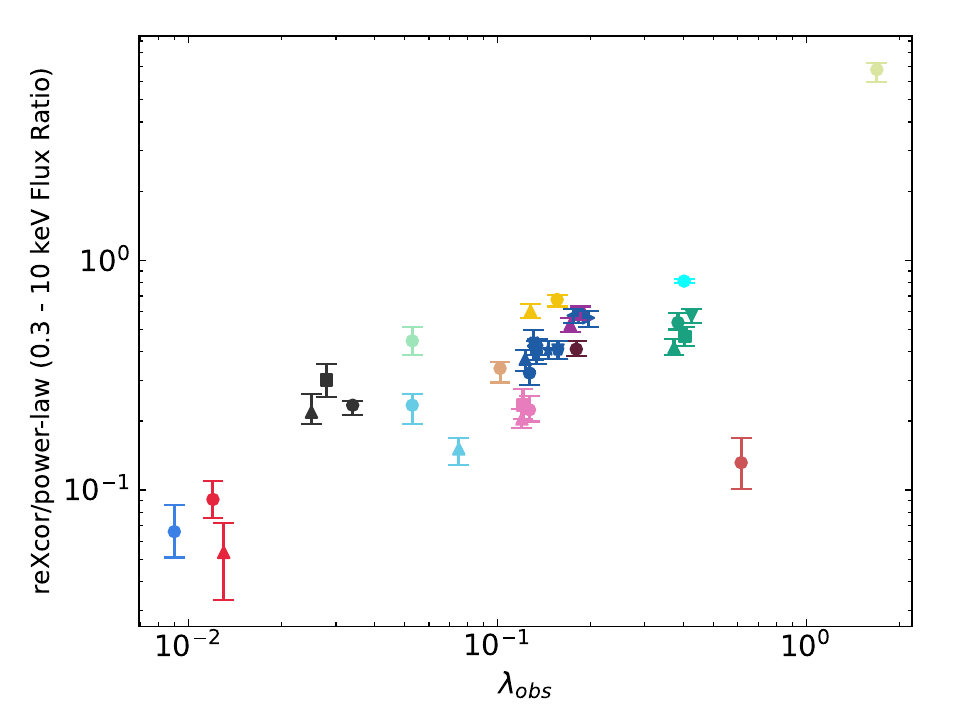}
    \caption{As in Figs.~\ref{fig:fx}--\ref{fig:fluxratio}, but now using the results of Table~\ref{table:bf2} for the 5 observations that give a negative $\Delta \chi^2$ when using the $\lambda=0.1$ \rexcor\ grids. The most significant changes in the \rexcor\ parameters occur in two of the lowest \lobs\ in our sample (MRK590 and ESO198-G24 [Obs. ID 0305370101]). The resulting statistical analysis is shown in Table~\ref{table:stats2}, which shows that the $f_X$-\lobs, $\Gamma$-\lobs\ and the flux ratio-\lobs\ correlations remain significant, but the `v'-shaped $\tau$-\lobs\ and $h_f$-\lobs\ relationships are retained only when omitting MRK590 and ESO198-G24 [Obs. ID 0305370101].}
    \label{fig:appC}
\end{figure*}
The most significant change in the \rexcor\ parameters occurs in the two observations with the lowest \lobs\ in the sample (MRK590 and ESO198-G24 [Obs. ID 0305370101]). In both cases, $h_f$ and $\tau$ fall to much lower values than was originally found with the $\lambda=0.01$ \rexcor\ grids. Interestingly, the parameters found in the other ESO198-G24 observation (Obs. ID 0067190101) are consistent with the prior fit, despite having essentially the same \lobs\ (Table~\ref{table:sample}). If this new fit for Obs. ID 0305370101 is a better representation of the coronal properties then it would imply that AGN coronae can be substantially different in the same object even at approximately the same \lobs.

The impact of these changes on our statistical analysis is shown in Table~\ref{table:stats2}.
\begin{table*}
    \centering
    \caption{As in Table~\ref{table:stats}, but now showing the results of the correlation analysis when the \rexcor\ results for the five observations with negative $\Delta \chi^2$ in Table~\ref{table:bf2} are used in the analysis (Fig.~\ref{fig:appC}). The significance of the three relationships with \lobs\ presented in the top half of the Table are relatively unaffected with this change. Evidence for the `v' shape transition in the $\tau$ and $h_f$ relationships with \lobs\ is lost after switching in the new values, but is recovered after removing the two low \lobs\ sources that exhibited the largest change in $h_f$ and $\tau$.  }
    \label{table:stats2}
    \begin{tabular}{l|c|c}
    Relationship with \lobs\ & Kendall's $\tau$ & $p$-value \\ \hline
    $f_X$ & $-0.56\pm 0.06$ & $2.8^{+31}_{-2.5}\times 10^{-6}$ \\
    $\Gamma$ & $0.24\pm 0.06$ & $0.050^{+0.097}_{-0.037}$ \\
    $(0.3-10)$~keV Flux ratio & $0.56\pm 0.03$ & $2.8^{+6.6}_{-2.0}\times 10^{-6}$ \\
    $(0.3-10)$~keV Flux ratio (no RE1034+396) & $0.53\pm 0.03$ & $1.1^{+2.6}_{-0.8}\times 10^{-5}$ \vspace{4mm} \\
     &  & Mean $p$-value \\
    $\tau$ and $h_f$ transition \lobs\ (i.e., $\lambda_{\mathrm{obs,t}}$) & Mean $p$-value & (no MRK590 and ESO198-G24 [Obs. ID 0305370101])\\ \hline
    $0.10$ & $0.32^{+0.26}_{-0.18}$ & $0.16^{+0.33}_{-0.13}$\\
    $0.12$ & $0.36\pm 0.14$ & $0.062^{+0.199}_{-0.049}$\\
    $0.15$ & $0.10^{+0.11}_{-0.06}$ & $0.007^{+0.024}_{-0.005}$\\
    $0.20$ & $0.16^{+0.14}_{-0.11}$ & $0.12^{+0.08}_{-0.09}$\\
    \end{tabular}
\end{table*}
Comparing these values to the ones in Table~\ref{table:stats} shows that the three correlations with \lobs\ described in Sect.~\ref{sect:res} are retained here at comparable levels of significance. However, the evidence for `v'-shaped relationships of $\tau$ and $h_f$ with \lobs\ is lost unless MRK590 and ESO198-G24 [Obs. ID 0305370101]) are omitted from the analysis. Therefore, we conclude that the results described in Sect.~\ref{sect:res} are robust to changing the assumed $\lambda$ in the \rexcor\ modeling, except at Eddington ratios $\la 0.01$. It is possible that at these accretion rates, the disc is transitioning away from an optically thick flow (driven, perhaps, by the unknown second parameter needed to explain the warm corona evolution), and the \rexcor\ model would no longer be applicable.

For completeness, we also perform the opposite experiment where the 27 observations best fit with the $\lambda=0.1$ \rexcor\ grids are re-fit using the $\lambda=0.01$ models. A large majority of the 27 observations are poorly fit with these higher density, less ionized models, with the median $\Delta \chi^2=+19.4$. Only 5 of the observations have a $\Delta \chi^2 < 6$, with just two of these giving a negative $\Delta \chi^2$: MRK279, Obs. ID 03022480501 ($\Delta \chi^2=-0.2$) and PG1116+215, Obs. ID 0554380301 ($\Delta \chi^2=-3.5$). In both these cases, $h_f$ and $f_X$ are at their maximum values, showing that the fit was searching for the most ionized model available in the grid. Therefore, it is likely that the slight improvement in the fit statistic in these two cases is not significant. The other three observations with $\Delta \chi^2 < 6$ includes another PG1116+215 spectrum (Obs. ID 0554380201), which showed the same behavior as the other observation just discussed, and two MRK590 observations (Obs. ID 0601390701 and 0601391001). In the two MRK590 observations, $h_f$ increased from $\approx 0.4$ (as found in the $\lambda=0.1$ \rexcor\ grids) to $h_f \approx 0.6-0.7$, with $f_X$ and $\tau$ largely unchanged. However, as the 8 other MRK590 observations are poorly fit with the high density \rexcor\ grid, we conclude that this source is best described using the $\lambda=0.1$ models.

\subsection{The coronal height, $h$}
\label{sub:height}
For a specific $\lambda$ (either $\lambda=0.1$ or $0.01$), there are \rexcor\ grids calculated for two specific heights of the hot, X-ray emitting corona: $h=5$ or $20$~$r_g$ \citep{xb22}. The purpose of these grids is not to make a measurement of $h$, per se, but to distinguish between a low lamppost ($h=5$) and a high lamppost (at $h=20$), since they lead to different illumination patterns on the accretion disc \citep[e.g.,][]{fk07,dauser13,ball17}. 

Table~\ref{table:bestfits} shows the value of $h$ for the \rexcor\ grid that gives the lowest $\chi^2$. In 13 of the 34 observations (those indicated with a $\dag$ by the Obs. ID) fits with the alternative coronal height have a $\Delta \chi^2 < 6$, which, based on the $\Delta \mathrm{BIC}$ criterion described above, means that one value of $h$ cannot be preferred over the other. In 9 out of the 13 observations, the warm corona parameters found with the alternative $h$ are consistent with the ones shown in Table~\ref{table:bestfits}. For UGC3973 (Obs. ID 040070301) the fit with $h=20$ gives $\tau=23.3^{+0.4}_{-3.3}$ (rather than $16.5\pm 0.3$), which is still consistent with the observed trends seen in Fig.~\ref{fig:tauandhf}. A value of $h=5$ when fitting observation 0601390701 of MRK509 pushes $h_f$ to $0.76$, $\tau$ to $16.8$ and $f_X$ down to $0.02$.  When $h=20$ observation 030248501 of MRK279 yields $\tau=23.8^{+5.6}_{-3.8}$, inconsistent with the value of $\tau=14^{+4.1}_{-1}$ found when $h=5$. This new value of $\tau$ would pull this observation away from the `v'-shape seen in Fig.~\ref{fig:tauandhf}, reducing its significance. However, all the other observations of MRK509 and MRK279 give consistent \rexcor\ parameters when fit with either coronal height. Although we cannot rule out variability intrinsic to the source, since all the observations of these objects have approximately the same \lobs, we expect that the best-fit values shown in Table~\ref{table:bestfits} remain the most likely results. 

Lastly, a coronal height of $h=5$ decreases both $\tau$ and $h_f$ for UGC3973 (Obs. ID 0400070401) to $10.6^{+1.7}_{-0.6p}$ and $0.43^{+0.14}_{-0.11}$, respectively. These values are similar to the ones found in Obs. ID 0400070201 of the same source, and are inconsistent with the trends with \lobs\ seen in Fig.~\ref{fig:tauandhf}. The $h=20$ and $h=5$ fits both give a reduced $\chi^2 \approx 1$ and so it is difficult to distinguish the two cases. However, this is the only observation where a difference in $h$ may plausibly increase the scatter in the $\tau$-\lobs\ and $h_f$-\lobs\ planes.

\subsection{The black hole spin, $a$}
\label{sub:spin}
The \rexcor\ grids provide different choices for the black hole spin, $a=0.9$ and $0.99$ \citep{xb22}. As with the corona height, the purpose is not to measure a black hole spin with these models, but to test if the data requires an maximal spin as compared to a black hole that is simply rapidly spinning. In the \rexcor\ model, a higher spin not only increases the relativistic blurring and light-bending effects, but it also increases the flux dissipated into the coronae (which are fractions of the dissipation rate $D(r,\lambda)$; Sect.~\ref{sect:rexcor}). This effect can be compensated for by reducing $f_X$, so there is a moderate degeneracy between $a$ and $f_X$ in the \rexcor\ model \citep{xb22}.  

Each AGN in the sample was fit with \rexcor\ grids using both values of $a$ with the one associated with lowest $\chi^2$ model listed in Table~\ref{table:bestfits}. If an AGN has more than one observation than the value of $a$ found in the first observation was used for subsequent fits. For four AGNs (HE1029-1401, IRASF12397+3333, MRK509 and NGC4593; indicated with a `*' by the Obs. ID in Table~\ref{table:bestfits}), assuming the alternative spin when fitting gives a $\Delta \chi^2 < 6$, indicating that either value of $a$ may be considered acceptable. In the cases of HE1029-1401, IRASF12397+3333 and NGC4593, the resulting values of $f_X$, $h_f$, $\Gamma$ and $\tau$ are all consistent irrespective of the choice of $a$. For MRK509, assuming $a=0.9$ increases $f_X$ by $\approx 0.02$ in each of the observations compared to the values listed in Table~\ref{table:bestfits} with the other parameters consistent with the previous results. According to Fig.~\ref{fig:fluxratio}, such a modest change in $f_X$ for the MRK509 observations would not alter the observed trend between $f_X$ and \lobs.
%

\bsp	
\label{lastpage}
\end{document}